\documentclass[epj]{svjour}
\newcommand{\beao}{\begin{eqnarray*}}
\newcommand{\eeao}{\end{eqnarray*}}
\newcommand{\be}{\begin{equation}}\newcommand{\ee}{\end{equation}}
\newcommand{\bea}{\begin{eqnarray}}
\newcommand{\eea}{\end{eqnarray}}
\newcommand{\beq}{\begin{eqnarray}}
\newcommand{\eeq}{\end{eqnarray}}
\newcommand{\tra}{\top}
\newcommand{\nn}{\nonumber}
\newcommand{\pa}{\partial}
\newcommand{\ep}{\epsilon}
\newcommand{\al}{\alpha}\newcommand{\la}{\lambda}
\newcommand{\Ref}[1]{(\ref{#1})}
\renewcommand{\ss}{\sinh(s)}\newcommand{\st}{\sinh(t)}
\newcommand{\cs}{\cosh(s)}\newcommand{\ct}{\cosh(t)}
\newcommand{\czs}{\cosh(2s)}\newcommand{\szs}{\sinh(2s)}
\renewcommand{\sp}{\sinh(s+t)}\newcommand{\sm}{\sinh(s-t)}
\newcommand{\cp}{\cosh(s+t)}\newcommand{\cm}{\cosh(s-t)}
\newcommand{\p}{{p}}\newcommand{\q}{{q}}
\newcommand{\Q}{{X}}\newcommand{\Qb}{{Y}}\newcommand{\Qt}{{Z}}
\usepackage{epsfig}

\begin{document}
\title{Gluon polarization tensor in color magnetic background}
\author{
M. Bordag\thanks{e-mail: Michael.Bordag@itp.uni-leipzig.de}
\inst{1}    \and
 V. Skalozub\thanks{e-mail: Skalozub@ff.dsu.dp.ua}\inst{2}}

\institute{University of Leipzig, Institute for Theoretical Physics,
 Augustusplatz 10/11, 04109 Leipzig, Germany  \and  Dnepropetrovsk National University, 49050 Dnepropetrovsk, Ukraine }

\date{version of \today}
\textwidth18cm\topmargin-1cm
\oddsidemargin-0.5cm\evensidemargin-1.2cm

\abstract{
In SU(2) gluodynamics we calculate the gluon polarization tensor in an Abelian homogeneous magnetic field in one-loop order in the Lorentz background field gauge. It turned out to be non transversal and consisting of ten tensor structures and corresponding form factors - four in color neutral and six in color charged sector. Seven  tensor structures are transversal, three are not. The non transversal parts are obtained  by explicit calculation.  We represent the form factors in terms of double parametric integrals which can be computed numerically. Some examples are provided and possible applications are discussed.
}
\maketitle
\section{Introduction}

The investigation of the deconfinement phase in QCD has attracted a considerable interest in the past years. The current understanding is that this phase is not simply a noninteracting gas of quarks and gluons as assumed earlier but a nontrivial state of these fields. In particular, from lattice calculations \cite{Agasian:2003yw} and from perturbative daisy resummations \cite{Skalozub:2002da,Skalozub:2004ab} it was found that a chromomagnetic field of order $gB\sim g^4T^2$, where $g$ is the gauge coupling, $B$ is the magnetic field strength and $T$ is the temperature, is likely to appear  spontaneously.  These results are of importance not only  for the QCD but also for the possibility to explain the origin of the cosmological magnetic field in the early universe \cite{Pollock}.

In the perturbative approach  because of the known infrared problems it is necessary to make resummations, on the daisy level, on the super daisy level and, if possible, beyond. In any case the knowledge of the gluon polarization tensor in a magnetic field, including also non zero temperature and an $A_0$-condensate, is necessary.

The calculation of the polarization tensor in an external magnetic field has a long history, especially in QED. However in QCD much less is known. This is perhaps because in QCD it does not have a direct physical significance (except for the case of finite temperature for which it had been investigated in detail for example in \cite{Kalashnikov:1982sc}). Particular results had been obtained for example in \cite{Enqvist:1994rm}, \cite{Starinets:1994vi}, \cite{Elmfors:1998dr}, \cite{Skalozub:2002da,Skalozub:2004ab}.
 Among them the most important was the observation that in the background of a magnetic field the "charged"  gluons acquire a magnetic mass proportional to  $g^2 (gB)^{1/2} T$ \cite{Skalozub:2000ay}, whereas the "neutral" ones corresponding to the "Abelian projection" remain massless \cite{Skalozub:2004ab}. In general, the question about the  magnetic mass and whether it can cure the tachyonic instability  is discussed.

However it is a common feature of all these calculation that the polarization tensor in a magnetic field had never been calculated in detail but only in some approximations and projections on physical states, i.e., on-shell. Nevertheless its structure off-shell is a indispensable prerequisite for higher order resummations like applying the second Legendre transform. Also it is of interest for the $W$-bosons in the electroweak theory and for the question how to define the gluon magnetic mass.

So in the present paper we investigate the gluon polarization tensor in SU(2) gluodynamics in detail off shell in a homogeneous color magnetic background at zero temperature. We use the standard Lorentz background field gauge. Our first finding is that the tensor is not transversal, $p_\mu\Pi_{\mu\nu}\ne0$.  This fact is insofar remarkable as it is not in line with the common expectations. In fact its transversality in the presence of a magnetic field had never been shown. We find the nontransversal contributions from direct calculation. The structure of the polarization tensor is determined by the weaker  condition $p_\mu\Pi_{\mu\nu}p_\nu=0$. There are four independent tensor structures $T^{(i)}_{\mu\nu}$ for the color neutral polarization tensor and six for the charged. To each tensor structure there is a corresponding form factor.

It should be mentioned that the color neutral component of the polarization tensor had been calculated repeatedly in Fujikawa gauge, \cite{Nielsen:1978rm}, \cite{Borisov:1984di}, \cite{Skalozub:2002da,Skalozub:2004ab}. In that case the structure is much simpler, especially because the tensor is transversal. In addition it has only three tensor structures and accordingly only three form factors instead of four in background gauge (see below). But in this gauge the color charged components   are much more complicated.

In the next section we introduce the necessary notations and review the basic formulas. In the sections 3 and 4 we derive the general structure and explicit formulas in the representation of parameter integrals for the neutral and for the charged polarization tensor. In section 5 we discuss the renormalization and perform as an example some  numerical calculations.

Throughout the paper we use latin letters $a,b,\dots=1,2,3$ for the color indexes and Greek letters $\la,\mu,\dots=1,\dots,4$ for the Lorentz indices.  Summation over doubly appearing indices  is assumed. All formulas are in the Euclidean formulation unless otherwise indicated. We put all constants including the coupling  equal to unity.

\section{Basic Notations}\label{bn}
In this section we collect the well known basic formulas for SU(2) gluodynamics to set up
the notations which we will use. We work in the Euclidean version basically to avoid
unnecessary signs and factors $i$. Dropping arguments and indices the Lagrangean is simply
\be\label{BL}{\cal L}=-\frac14 F_{\mu\nu}^2-\frac{1}{2\xi}(\pa_\mu A_\mu)^2+\overline{\eta} \ \pa_\mu D_\mu \eta,
\ee
where $\xi$ is the gauge parameter and $\eta$ is the ghost field.
The action is $S=\int dx \ L$ and the generating functional of the
Green functions is $Z=\int DA \exp(S)$. In the following we divide
the gauge field $A_\mu^a(x)$ into background field $B_\mu^a(x)$
and quantum fluctutions $Q_\mu^a(x)$,
\be\label{ABQ}A_\mu^a(x)=B_\mu^a(x)+Q_\mu^a(x).
\ee
The covariant derivative depending on a field $A$ is
\be\label{covder}D_\mu^{ab}[A]=\frac{\pa}{\pa x^\mu}\delta^{ab}+\ep^{acb}A_\mu^c(x)
\ee
and the field strength is
\be\label{fs}F_{\mu\nu}^a[A]=\frac{\pa}{\pa x^\mu}A_\nu^a(x)-\frac{\pa}{\pa x^\nu}A_\mu^a(x)+\ep^{abc}A_\mu^b(x)A_\nu^c(x)
\ee
and
\be\label{komm}\left[D_\mu[A],D_\nu[A]\right]^{ab}=\ep^{acb}F_{\mu\nu}^c[A]
\ee
holds. For the field splitted  into background and quantum we note
\bea\label{F}F_{\mu\nu}^a[B+Q]&=&F_{\mu\nu}^a[B]+D_\mu^{ab}[B]Q_\nu^b(x)
-D_\nu^{ab}[B]Q_\mu^b(x) \nn \\ &&
+\ep^{abc}Q_\mu^b(x)Q_\nu^c(x). \eea
The square of it is
\bea\label{FBQ}-\frac14 \left(F_{\mu\nu}^a[B+Q]\right)^2&=&
-\frac14 \left(F_{\mu\nu}^a[B]\right)^2+Q_\nu^aD_\mu^{ab}[B]F_{\mu\nu}^b[B]\nn \\ &&
-\frac12 Q_\mu^a K_{\mu\nu}^{ab}Q_\nu^b
+{\cal M}_3+{\cal M}_4.
\eea
The second term in the r.h.s. is linear in the quantum field and disappears if the background fulfills its equation of motion which will hold in our case of a constant background. The third term is quadratic in $Q_\mu^a$ and it defines the 'free part' with the kernel
\be\label{K}K_{\mu\nu}^{ab}=-
\delta_{\mu\nu}D_\rho^{ac}[B]D_\rho^{cb}[B]+D_\mu^{ac}[B]
D_\nu^{cb}[B] -2\ep^{acb}F_{\mu\nu}^c[B]. \ee
The interaction of the quantum field  is represented by the  vertex factors
\bea\label{vfM}{\cal M}_3&=&-\ep^{abc}\left(D_\mu^{ad}Q_\nu^d\right)Q_\mu^bQ_\nu^c ,\nn \\
{\cal M}_4&=&-\frac14
Q_{\mu}^{a}Q_{\nu}^{a}Q_{\mu}^{b}Q_{\nu}^{b} + \frac14
Q_{\mu}^{a}Q_{\nu}^{b}Q_{\mu}^{a}Q_{\nu}^{b} . \eea
The complete Lagrangean
\be\label{comLa}{\cal L}=-\frac14 \left(F_{\mu\nu}^a[B+Q]\right)^2+{\cal L}_{\rm gf}+{\cal L}_{\rm gh}
\ee
consists of \Ref{FBQ}, the gauge fixing term (in the following we put $\xi=1$),
\be\label{Lgf}{\cal L}_{\rm gf}=-\frac{1}{2\xi}\left(D_\mu^a[B] Q_\mu^a\right)^2=
\frac{1}{2\xi}Q_\mu^a D_\mu^{ac}[B]D_\nu^{cb}[B]Q_\nu^b,
\ee
and the ghost term
\be\label{Lgh}{\cal L}_{\rm
gh}=\overline{\eta}^aD_\mu^{ac}[B]\left(D_\mu^{cb}[B]+\ep^{cdb}Q_\mu^d\right)\eta^b.
\ee
These formulas are valid for an arbitrary background field.  Now we turn to the specific background of an Abelian homogeneous magnetic field of strength $B$ which is oriented along the third axis in both, color and configuration space. An explicit representation of its vector potential is
\be\label{Bexpl}B_\mu^a(x)=\delta^{a3}\delta_{\mu 1}x_2 \ B
\ee
and the corresponding field strength is
\be\label{}F_{ij}^a=\delta^{a3}F_{ij}=B\ep^{3ij},\ee
where only the spatial components ($i,j=1,2,3$) are non\-zero.
Once the background is chosen Abelian it is useful to turn into the so called charged basis,
\bea\label{}W^\pm_\mu&=&\frac{1}{\sqrt{2}}\left(Q_\mu^1\pm i Q_\mu^2\right) \nn \\
Q_\mu&=&Q_\mu^3
\eea
with the interpretation of $W^\pm_\mu$ as color charged fields and $Q_\mu$ as color neutral field. This is in parallel to electrically charged and neutral fields. Note also that $Q_\mu$ is real while $W^\pm_\mu$ are complex conjugated one to the other. In the following we will omit the word color when speaking about charged and neutral objects.
The same transformation is done for the ghosts,
\bea\label{}\eta^\pm_\mu&=&\frac{1}{\sqrt{2}}\left(\eta_\mu^1\pm i \eta_\mu^2\right) ,\nn \\
\eta_\mu&=&\eta_\mu^3.
\eea
A summation over the color indices turns into
\be\label{}Q^a_\mu Q^a_\nu=Q_\mu Q_\nu+W^+_\mu W^-_\nu+W^-_\mu W^+_\nu .
\ee
All appearing quantities have to be transformed into that basis.
For the covariant derivative we get
\be\label{}\begin{array}{rclrclll}D^{33}_\mu&=&\pa_\mu,&
D^{-+}_\mu&=&\pa_\mu-iB_\mu&\equiv& D_\mu \\ [8pt]
&&& D^{+-}_\mu&=&\pa_\mu+iB_\mu  &\equiv& D^*_\mu \end{array}
\ee
where $D^*_\mu$ is the complex conjugated to $D_\mu$. Starting from here we do not need any longer to indicate the arguments in the covariant derivatives.

Before proceeding with writing down the remaining formulas in the
charged basis it is useful to turn into momentum representation.
This can be done in a standard way by the formal rules. It remains to define  the signs in the exponential factors. We adopt the notation
\be\label{}Q\sim e^{-ikx}, \qquad W^-\sim e^{-ipx}, \qquad W^+\sim
e^{ip'x}. \ee
In all following calculation the momentum $k$ will denote the momentum of a neutral line and the momenta $p$ and $p'$ that of the charged lines whereby $k$ and $p$ are incoming and $p'$ is outgoing. In these notations the covariant derivative $D_\mu$ acts on a $W^-_\mu$ and turns into
\be\label{defp}D_\mu=-i(i\pa_\mu+B_\mu)\equiv -i p_\mu .
\ee
Note that the components of the momentum $p_\mu$ do not commute,
\be\label{noncp}[p_\mu,p_\nu]=iBF_{\mu\nu}.
\ee
In these notations the quadratic term of the action turns into
\bea&&\label{q2}
-\frac12 Q_\mu^a K_{\mu\nu}^{ab}Q_\nu^b= \\ &&
\frac12 Q_\mu K_{\mu\nu}^{33} Q_\nu+\frac12 W^+_\mu K_{\mu\nu}^{-+}W^-_\nu+
\frac12 W^-_\mu K_{\mu\nu}^{+-}W^+_\nu \nn
\eea
with
\be\label{Kk} K_{\mu\nu}^{33}\equiv K_{\mu\nu}(k)=\delta_{\mu\nu}k^2-k_\mu k_\nu
\ee
and
\be\label{Kp}K_{\mu\nu}^{-+}\equiv
K_{\mu\nu}(p)=\delta_{\mu\nu}p^2-p_\mu p_\nu + 2iBF_{\mu\nu}. \ee
We use the arguments $k$ and $p$ instead of the indices to
indicate to which line a $K_{\mu\nu}$ belongs. The third term in
the r.h.s. of Eq.\Ref{q2} is the same as the second one due to the
complex conjugation rules. In the Feynman rules
$(-K_{\mu\nu}^{33})^{-1}$ is the line for neutral gluons and is
denoted by a wavy line and $(-K_{\mu\nu}^{-+})^{-1}$ is the line
for charged gluons and is denoted by a directed solid line. We
remark that these lines represent propagators in the background of
the magnetic field. Frequently they are denoted by thick or double
lines. Because we have in this paper no other lines the notation
with ordinary (thin) lines is unique.

For later use we introduce here the set of eigenstates for the
operators  \Ref{Kk} and \Ref{Kp}. For the color neutral states we
can take exactly the same polarizations $\mid k,s>$ as known from
electrodynamics,
\be\label{qedstates}
\begin{array}{rclrcl}
\mid k,1>_\mu&=&\frac{1}{k_{||}}\left(\begin{array}{c} -k_2\\k_1\\0\\0\end{array}\right)_\mu,&
\mid k,2>_\mu&=&\frac{1}{k k_{||}}\left(\begin{array}{c} k_1 k_3\\k_2 k_3\\-k_{\small ||}^2\\0\end{array}\right)_\mu , \\
\mid k,3>_\mu&=&\frac{1}{k}\left(\begin{array}{c} k_1\\k_2\\k_3\\0\end{array}\right)_\mu,&
\mid k,4>_\mu&=&\left(\begin{array}{c} 0\\0\\0\\1\end{array}\right)_\mu
\end{array}
\ee with $k_{||}=\sqrt{k_1^2+k_2^2}$,
$k=\sqrt{k_1^2+k_2^2+k_3^2}$. Here the polarizations  $s=1,2$
describe the two transversal gluons ($k_\mu\mid k,s=1,2>_\mu=0$),
$s=3$ is the longitudinal one and $s=4$ after rotation into
Minkowski space becomes the time like one. For the transversal
gluons
\be\label{eomK} K_{\mu\nu}(k)  \ \mid k,s=1,2>_\nu=(k_4^2+k^2) \ \mid k,s=1,2>_\mu
\ee
holds.

For the color charged gluons we denote the states by $\mid
p_{||},n,s>_\mu$ where  the integer $n$ corresponds to the Landau
levels and  $p_{||}=\{p_3,p_4\}$ are the momenta in the direction
parallel to the magnetic field. In opposite to the color neutral
gluons whose momenta are completely continuous and whose wave
functions are plane waves, for the color charged gluon we have a
spectrum similar to an electrically  charged particle in a
magnetic field, contineous momenta in the directions parallel to
the magnetic field (in our case $\mu=3$ and $\mu=4$) and the
integer $n$ which corresponds to the Landau levels. In fact there
is one more quantum number which for example in the representation
\Ref{Bexpl} describes the $x$-coordinate of the centers of the
cyclotron orbits. We drop it because nothing we are interested in
dependence on it.

In order to describe the states $\mid p_{||},n,s>_\mu$ it is
useful to turn the  coordinate system in the plane perpendicular
to the magnetic field, i.e., in the $(p_1,p_2)$-plane, according
to
\be\label{Bma}
p_\mu=B_{\mu\alpha}\ p_\alpha \quad \mbox{with} \quad B_{\mu\alpha}=\left(\begin{array}{cc}\frac{1}{\sqrt{2}}
\left(\begin{array}{cc}1&1\\i&-i\end{array}\right) & 0\\ 0 & \left(\begin{array}{cc}1&0\\0&1\end{array}\right) \end{array}\right)_{\mu\alpha} ,
\ee
where $\mu,\nu$ are the Lorentz indices before and $\alpha,\beta$
after the rotation. The rotated momentum is
\be\label{palpha}p_\alpha=\left(\begin{array}{c}ia^\dagger\\-ia\\p_3\\p_4\end{array}
\right)_{\alpha},
\ee
where $a=\frac{1}{\sqrt{2}}(ip_1-p_2)$ and
$a^\dagger=\frac{1}{\sqrt{2}}(-ip_1-p_2)$  are the harmonic
oscillator  ladder operators, $[a,a^\dagger]=1$. The field
strength is diagonal in this representation,
\be\label{Fsigma3}
F_{\alpha\beta}=\left(\begin{array}{cccc}1&0&0&0\\0&-1&0&0\\0&0&0&0\\0&0&0&0
\end{array}\right)_{\alpha\beta} .\ee
Now we denote by $\mid n>$ the standard harmonic oscillator eigenstates, $\mid n>=\frac{1}{\sqrt{n!}}\left(a^\dagger\right)^n\mid 0>$. Together with a second harmonic oscillator which corresponds to the quantum number which we dropped they describe the coordinate dependence of the states in the plane perpendicular to the magnetic field. A fairly complete representation of these states (and of the tree level gluon propagator) we found in  \cite{Kaiser:1987cc}.

In these notations the states of the charged gluons are
\be\label{chstates}
\begin{array}{rclrcl}
\mid p_{||},n,1\rangle_\alpha &=&\frac{1}{h\sqrt{n(n+1)}}\left(\begin{array}{c} a^\dagger n \\ a (n+1)\\0\\0\end{array}\right)_\alpha\mid n\rangle ,\\
\mid p_{||},n,2\rangle_\alpha &=&\frac{1}{hk}\left(\begin{array}{c} ia^\dagger l_3 \\ -ia l_3\\-h^2\\0\end{array}\right)_\alpha\mid n\rangle , \\
\mid p_{||},n,3\rangle_\alpha &=&\frac{1}{k}\left(\begin{array}{c} ia^\dagger  \\ -ia \\l_3\\0\end{array}\right)_\alpha\mid n\rangle ,  \\
\mid p_{||},n,4\rangle_\alpha &=&\left(\begin{array}{c} 0\\0\\0\\1\end{array}\right)_\alpha\mid n\rangle , \end{array}
\ee
where we introduced the notations $l^2=p_3^2+p_4^2$, $h^2=(2n+1)B$, $k=\sqrt{p_3^2+h^2}$ and
\be\label{treelevel}K_{\al\beta}(p) \ \mid p_{||},n,s\rangle_\beta = (l^2+h^2 )\mid p_{||},n,s\rangle_\al
\qquad (s=1,2)\ee
holds.

The tachyonic mode is
\be\label{tmode}\mid p_{||},-1,1\rangle_\al=\left(\begin{array}{c}1\\0\\0\\0\end{array}\right)\mid 0\rangle .
\ee
It has $h^2=-B$. In order to get it from \Ref{chstates} one needs
to rewrite $ \mid p_{||},n,1\rangle_\alpha =
\frac{1}{h}\left(\begin{array}{c}\sqrt{n}\mid n+1\rangle
\\\sqrt{n+1}\mid n-1\rangle\\0\\0\end{array}\right)_\al $ using
\\$a^\dagger\mid n\rangle=\sqrt{n+1}\mid n+1\rangle$ etc. There is
no state  with $n=-1$ for $s=2$. Also, there is no state with
$n=0$ for $s=1$, but   one for $n=0$, $s=2$,
\be\label{0mode} \mid p_{||},0,2\rangle_\alpha =
\frac{1}{k}\left(\begin{array}{c}ip_3
\\0\\\-B\\0\end{array}\right)_\al \mid0\rangle. \ee These are the
two lowest states, they are singlets. All higher states have
$n=1,2,\dots$,   $s=1,2$ and they are doublets. The states with
$s=1,2$ are transversal, $p_\mu \mid p_{||},n,s\rangle_\mu =0$
($s=1,2$), i.e., they fulfil the subsidiary condition. The states
\Ref{chstates}, \Ref{tmode}, \Ref{0mode} as well as
\Ref{qedstates} form each a basis of polarizations.

For completeness we note that after turning to Min\-kowski space
the   one particle energies $p_0^2=-p_4^2$ of these states are
$p_3^2+h^2=p_3^2+B(2n+1)$ with $n=-1$ for the tachyonic mode
\Ref{tmode}.  Further we note that for vanishing magnetic field
these states turn into \Ref{qedstates}.

\begin{figure}\unitlength0.5cm
\hspace{3cm}\begin{picture}(10,8)\label{figlevels}
\put(-3,5){\mbox{a.)}}
\put(0,1){\line(1,0){2}}\put(0,3){\line(1,0){2}}\put(0,5){\line(1,0){2}}
\put(4,-1){\line(1,0){2}}\put(4,1){\line(1,0){2}}
\put(4,3){\line(1,0){2}}\put(4,2.9){\line(1,0){2}}
\put(4,5){\line(1,0){2}}\put(4,4.94){\line(1,0){2}}
\put(2,1){\line(1,-1){2.05}}\put(2,3){\line(1,-1){2.05}}\put(2,5){\line(1,-1){2.05}}
\put(2,1){\line(1,1){2.05}}\put(2,3){\line(1,1){2.05}}
\put(2,5){\line(1,1){0.8}}\put(4,5){\line(-1,1){0.8}}
\multiput(-0,0)(0.9,0){8}{\line(1,0){0.4}}
\put(-1,-0.3){\mbox{\small 0}}\put(-1,0.7){\mbox{\small 1}}\put(-1,2.7){\mbox{\small 3}}\put(-1,4.7){\mbox{\small 5}}
\put(0,-2.5){\mbox{$E_n^2=B(2n+1\pm 1)$}}
\end{picture}
\vspace*{1cm}

\hspace*{3cm}\begin{picture}(10,8)
\put(-3,5){\mbox{b.)}}
\put(0,1){\line(1,0){2}}\put(0,3){\line(1,0){2}}\put(0,5){\line(1,0){2}}

\put(3,0.2){\line(1,0){2}}
\put(3,2){\line(1,0){2}}\put(3,1.95){\line(1,0){2}}
\put(3,4){\line(1,0){2}}\put(3,3.9){\line(1,0){2}}

\put(2,1){\line(1,-1){0.9}}\put(2,3){\line(1,-1){0.9}}\put(2,5){\line(1,-1){0.9}}

\put(2,1){\line(1,1){0.9}}\put(2,3){\line(1,1){0.9}}
\put(2,5){\line(1,1){0.8}}
\put(-1,-0.3){\mbox{\small 0}}\put(-1,0.7){\mbox{\small 1}}\put(-1,2.7){\mbox{\small 3}}\put(-1,4.7){\mbox{\small 5}}
\multiput(-0,0)(0.9,0){7}{\line(1,0){0.4}}
\put(0,-2.){\mbox{$E_n^2=B(2n+1\pm \frac12)$}}
\end{picture}

\vspace{1.5cm}\caption{The level scheme in a magnetic field, a.)
for a color charged  gluon, b.) for a spinor. Here $n$ counts the
Landau levels and it is different from the $n$ in the states
\Ref{chstates}.}
\end{figure}
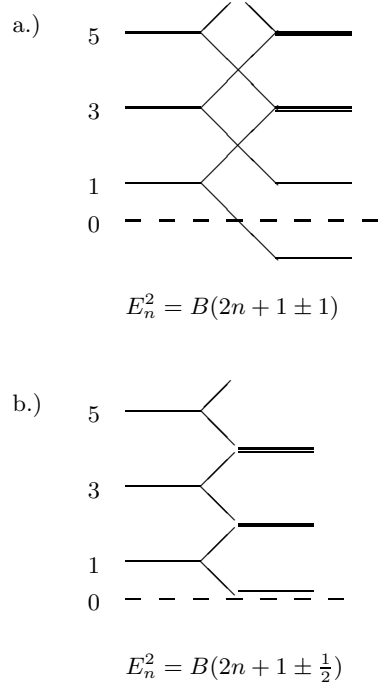

A more common representation would be $p_3^2+B(2n+1\pm\sigma)$
with $n=0,1,\dots$ for the Landau levels of a scalar field and
with the spin  projection $\sigma=\pm 1$ (for a spinor field
$\sigma=\pm \frac12$). A scheme of the levels in this
representation is shown in Fig.\ref{figlevels} where for
comparizon the spinor case is shown too. To derive this
representation in the literature the following discussion can be
found. Taking into account the subsidiary condition, $p_\mu\mid
n,s\rangle_\mu=0$, the equation \Ref{eomK} can be simplified by
dropping $p_\mu p_\nu$ in \Ref{Kp}. After that the remaining two
operators, $\delta_{\mu\nu}p^2$ and $2iF_{\mu\nu}$, commute and
their common eigenvectors can be taken proportional to
\be \mid \sigma=1\rangle\sim\left(\begin{array}{c}1\\0\\0\\0\end{array}\right) \qquad \mid \sigma=-1\rangle\sim\left(\begin{array}{c}0\\1\\0\\0\end{array}\right)
\ee
with the interpretation of spin up and spin down state. However,
these  two states cannot fulfil the subsidary condition. For that
reason we prefer the states \Ref{chstates}.

\begin{figure}[htbp]
\begin{center}
\ \psfig{height=3cm,file=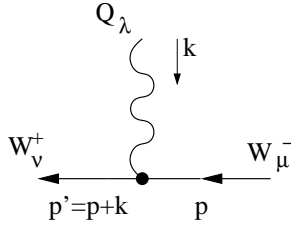}
\end{center}
\caption{Notations for the vertex ${\cal M}_3$}
\label{figure:Gamma3}
\end{figure}

After discussing the free part of the Lagrangean \Ref{comLa},
\Ref{FBQ},  in the 'charged basis' we note that for the vertex
factor ${\cal M}_3$ in  \Ref{vfM} we get
\be\label{M3}{\cal M}_3=W^-_\mu\Gamma_{\mu\nu\la}W^+_\nu Q_\la
\ee
with
\be\label{Gamma}\Gamma_{\mu\nu\la}=g_{\mu\nu}(k-2p)_\la+g_{\la\mu}(p+k)_\nu+g_{\la\nu}(p-2k)_\mu .
\ee
The notations are shown in Fig.\ref{figure:Gamma3}. Similar
expressions appear for the  vertexes involving ghosts. It should
be remarked that all graphs and combinatorical factors are exactly
the same as in the well known case without magnetic field. On this
level the only difference is in the meaning of the momentum
$p_\mu$ which   in our case depends on the background magnetic
field, see Eq.\Ref{defp}.

\unitlength1cm
\begin{figure}[htbp]\begin{picture}(10,2.9)
\put(0.6,1){ \psfig{height=1.7cm,file=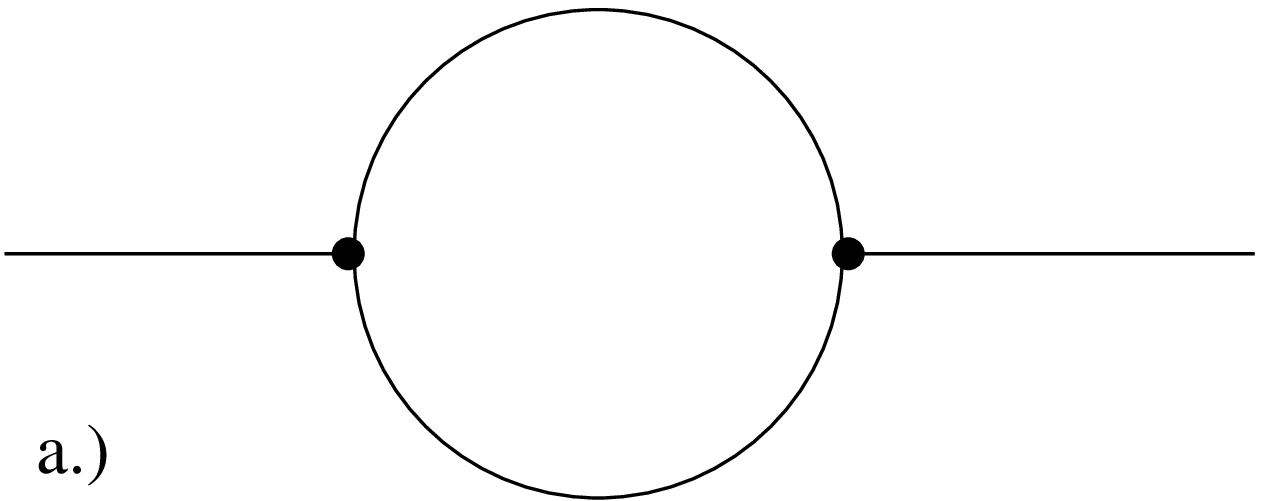}}
\put(5,0.21){ \psfig{height=2cm,file=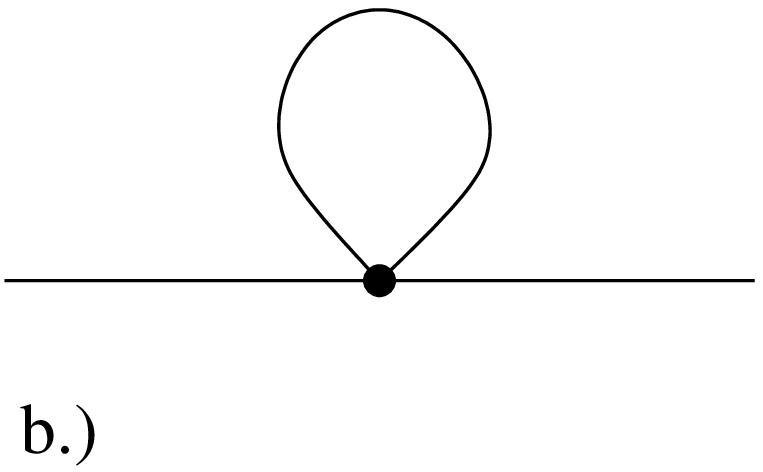}}
\end{picture}
\caption{a.) Basic graph for polarization tensor, b.) Graph with
one  vertex and a closed line} \label{figure:G34}
\end{figure}

The polarization tensor consists basically of the graphs shown in
Fig.  \ref{figure:G34} where gluon and ghost lines must be
inserted. For the calculaion we use the following strategy. We
drop all contributions which do not depend on an external
momentum.
This includes for instance the graph b.) in Fig. \ref{figure:G34}
with one vertex and a closed line over it. Also we will later
integrate by parts in the parameter integrals and we will  drop
all boundary contributions. The justification for this procedure
is that all these contributions in the final result must cancel.
This had been observed explicitely in the calculation of the
neutral polarization tensor in Fijikawa gauge. The general
argument is that for homogenuity reasons the polarization tensor
depends on the external momenta $p$ only divided by the field
strength, $p^2/B$. But for $B\to 0$ we have a normalization
condition which removes all freedom in adding a constant. Another
argument follows from the dispersion relations representing the
polarization tensor in terms of its jump across the cut it has.
But all contributions which we drop depend on the momentum only
polynomially hence they do not contribute to the jump.

\section{The Neutral Polarization Tensor}
The neutral polarization tensor is shown in Fig.
\ref{figure:Pineu}.  It is denoted by $\Pi_{\la\la'}(k)$ where the
argument $k$ indicates that it depends on an ordinary momentum
which is a number in opposite to the momentum $p$. In the next
subsection we discuss its general tensor structure and in the
subsequent subsections we perform actually its calculation.

\subsection{Operator Structures}\label{sectopstr}
Since the polarization tensor is not transversal in a magnetic
field, i.e., since $k_\la\Pi_{\la\la'}(k)=0$ does not hold, we are
left with the weaker condition
\be\label{doubletransv} k_\la \ \Pi_{\la\la'}(k) \ k_{\la'}=0 . \ee
It can be combined with the remaining Lorentz symmetry which
results in  a dependence of $\Pi_{\la\la'}(k)$ on  two vectors,
$l_\la$ and $h_\la$, and on the magnetic field.

We use the notations \bea&&\label{vec}
l_\mu=\left(\begin{array}{c}0\\0\\k_3\\k_4\end{array}\right),~
h_\mu=\left(\begin{array}{c}k_1\\k_2\\0\\0\end{array}\right), ~
d_\mu=\left(\begin{array}{c}k_2\\-k_1\\0\\0\end{array}\right),~
\nn\\&&
F_{\mu\lambda}=\left(\begin{array}{cccc}0&1&0&0\\-1&0&0&0\\0&0&0&0\\0&0&0&0\end{array}
\right).\eea The third vector is $d_\mu\equiv F_{\mu\nu}k_\nu$
with the field strength  of the magnetic field which is connected
with the non commuting components of the momentum of a charged
field by means of $[p_1,p_2]=iF_{12}$. Here and in the rest of the
section the magnetic field is put equal to unity. For the vectors
$k_\la=l_\la+h_\la$ holds. We note that the components of the
momentum $k_\mu$ are numbers which commute. The same holds in this
section for $h_\mu$.

The general structure of $\Pi_{\la\la'}(k)$ allowed by
\Ref{doubletransv} and the vectors $l_\la$ and $h_\la$ is
determined by the set of tensor structures
\begin{eqnarray} \label{Tn}
T^{(1)}_{\la\la'}&=& l^2 \delta^{||}_{\la\la'}-l_\la l_{\la'}\nn \\
T^{(2)}_{\la\la'}&=&h^2\delta^\perp_{\la\la'}
-h_\la h_{\la'}=d_\la d_{\la'}    \nn \\
T^{(3)}_{\la\la'}&=& h^2 \delta^{||}_{\la\la'}+
l^2 \delta^\perp_{\la\la'} -l_\la h_{\la'}-h_\la l_{\la'}\nn \\
T^{(4)}_{\la\la'}&=& i(l_\la d_{\la'}-d_\la l_{\la'} )
+il^2F_{\la\la'}\nn \\
T^{(5)}_{\la\la'}&=&h^2 \delta^{||}_{\la\la'}-l^2 \delta^\perp_{\la\la'} \nn \\
T^{(6)}_{\la\la'}&=&iF_{\la\la'}
\end{eqnarray}
together with the identity
$d_{\la}h_{\la'}-h_{\la}d_{\la'}=h^2F_{\la\la'}$. Further  we
introduced the notations
$\delta^\perp_{\mu\lambda}=\mbox{diag}(1,1,0,0)$ and
$\delta^{||}_{\mu\lambda}=\mbox{diag}(0,0,1,1)$. The first four
operators are transversal, $k_\la
T^{(i)}_{\la\la'}=T^{(i)}_{\la\la'}k_{\la'}=0$ with $i=1,2,3,4$,
the last two fulfil \Ref{doubletransv} only. The sum of the first
three operators is just the transversal part of the kernel of the
quadratic part of the action, Eq. \Ref{Kk},
\be\label{sumTip}
T^{(1)}_{\la\la'}+T^{(2)}_{\la\la'}+T^{(3)}_{\la\la'}=K_{\la\la'}(k).
\ee

Below we will be concerned with structures as they appear from the calculation of graphs.
Here we collect some formulas which will be helpful to organize the results in terms of form
factors. In general, knowing the operators \Ref{Tn} which may appear, the polarization tensor
can be represented in the form
\be\label{exp} \Pi_{\la\la'}(k)=\sum_{i=1}^6 \ \Pi^{(i)}(k) \  T^{(i)}_{\la\la'} \ ,\ee
where the 'form factors', $\Pi^{(i)}(k)$, depend on $l^2$ and on
$h^2$ only. Actually,  since the polarization tensor
$\Pi_{\la\la'}(k)$ is real and symmetric in its indices, the form
factors $\Pi^{(4)}(k)$ and $\Pi^{(6)}(k)$ are zero.

In the course of calculation, from the graph (Fig. \ref{figure:Pineu}) contributions   will
appear which have the following form. First,
\be\label{ts1} \Pi_{\la\la'}=P_\la P^\top_{\la'}
+a\delta^{||}_{\la\la'}+b\delta^{\perp}_{\la\la'}+icF_{\la\la'},
\ee
where $P_\la$ is given in terms of the vectors \Ref{vec},
\be \label{Pla} P_{\la}=r  \ l_\la+\al  \  id_\la +\beta  \ h_\la.\ee
The transposition in $P^\top_{\la'}$ changes the sign of $d_{\la'}$, $P_{\la'}^\top=r  \ l_\la-\al  \  id_\la +\beta  \ h_\la$. The expression in Eq.\Ref{exp} fulfills \Ref{doubletransv} if $\left(rl^2+\beta h^2\right)^2+al^2+bh^2=0$ holds. In that case it can be represented in terms of form factors according to
\bea\label{rep1}&&\Pi_{\la\la'}=-r^2T^{(1)}_{\la\la'}+(\al^2-\beta^2)T^{(2)}_{\la\la'}-r\beta T^{(3)}_{\la\la'}-r\al T^{(4)}_{\la\la'}\nn\\&&
+\frac{r(rl^2+\beta h^2)+a}{h^2}T^{(5)}_{\la\la'}+(ral^2+\al\beta h^2+c)T^{(6)}_{\la\la'}
.\eea
A second type of expressions appears which has a  slightly more complicated form,
\be\label{ts2} \Pi_{\la\la'}=P_\la Q_{\la'}+Q^\top_\la P^\top_{\la'} +a \delta^{||}_{\la\la'}+b \delta^\perp_{\la\la'}+icF_{\la\la'}\ee
with $P_\la$ from \Ref{Pla} and
\be\label{Qla} Q_\la=s \ l_\la+\gamma \ id_\la+ \delta \ \ h_\la.\ee
If for \Ref{ts2} the condition \Ref{doubletransv} is fulfilled   $(a+2rsl^2)l^2+(b+2\beta\delta h^2)h^2+(r\delta+s\beta)2l^2h^2=0$ must hold. In that case  the representation in terms of form factors is
\bea\label{rep2}\Pi_{\la\la'}&=&-2rsT^{(1)}_{\la\la'}-2(\beta\delta+\al\gamma)T^{(2)}_{\la\la'}
-(r\delta+s\beta) T^{(3)}_{\la\la'}\nn\\&&+(r\gamma-s\al) T^{(4)}_{\la\la'}
+\left((a+2rsl^2)\frac{1}{h^2}+r\delta+s\beta\right) T^{(5)}_{\la\la'}
\nn\\&&
+\left(c-(r\gamma-s\alpha)l^2+(\al\delta-\beta\gamma)h^2\right)T^{(6)}_{\la\la'}
.\eea

\subsection{Calculation of the Neutral Polarization Tensor}\label{sect2}
\begin{figure}[htbp]\begin{picture}(10,3.2)
\put(1,0.19){ \psfig{height=3cm,file=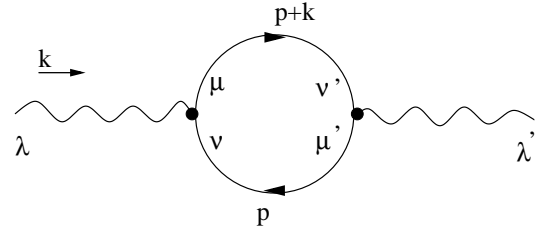}}
\end{picture}
\caption{The neutral polarization tensor}
\label{figure:Pineu}
\end{figure}

The neutral polarization tensor has the following representation
in momentum space  (see Fig. \ref{figure:Pineu})
\bea\label{NPT1}
&&\Pi_{\la\la'}(k)=\Gamma_{\mu\nu\la}G_{\mu\mu'}(p)\Gamma_{\mu'\nu'\la'}
G_{\nu'\nu}(p-k)\\&&
-p_{\la}G(p)(p-k)_{\la'}G(p-k)-(p-k)_{\la}G(p)p_{\la'}G(p-k),\nn
\eea where the integration over the momentum $p$ is assumed. The
second line is the contribution from the ghost loop.

The vertex factor is
\be\label{vfbg}\Gamma_{\mu\nu\la}=g_{\mu\nu}(k-2p)_{\la}+g_{\la\mu}(p+k)_{\nu}+
g_{\la\nu}(p-2k)_{\mu}.
\ee
For a convenient grouping of terms it is useful to rearrange it,
\be\label{vertexfactor} \begin{array}{lcccccc}
\Gamma_{\mu\nu\la}= \\
\underbrace{g_{\mu\nu}(k-2p)_{\la}}&+&
\underbrace{2\left(g_{\la\mu}k_\nu-g_{\la\nu}k_\mu\right)}&+&
\underbrace{g_{\la\mu}(p-k)_\nu+g_{\la\nu}p_\mu} , \\ [10pt]\equiv
\Gamma_{\mu\nu\la}^{(1)}&+&\Gamma_{\mu\nu\la}^{(2)}&+&\Gamma_{\mu\nu\la}^{(3)},
\end{array}
\ee
where in the last line a subdivision into three parts is done.

The propagators are given by
\bea\label{prop0} &&G(p)=\frac{1}{p^2}=\int_0^\infty ds \ e^{-sp^2}, \nn \\ &&
G(p-k)=\frac{1}{(p-k)^2}=\int_0^\infty dt \ e^{-t(p-k)^2}  \eea
 for the scalar lines
and by \bea\label{prop} G_{\la \la'} (p)&=&\left(\frac{1}{p^2+2iF}\right)_{\la\la'}=\int_0^\infty ds \
e^{-sp^2}E^{-s}_{\la\la'}  ,\nn\\
G_{\la \la'}
(p-k)&=&\left(\frac{1}{(p-k)^2+2iF}\right)_{\la\la'}\\&&~~~~~~~~~=\int_0^\infty dt \ e^{-t(p-k)^2}E^{-t}_{\la\la'} \nn \eea for the
vector lines (in the Feynman gauge, $\xi = 1)$ with
\bea\label{E}E^s_{\la\la'} &\equiv& \left(e^{2isF}\right)_{\la\la'}  \nn \\&=&
\delta^{||}_{\la\la'}+iF_{\la\la'}\sinh(2s)+\delta^\perp_{\la\la'} \cosh(2s) .
\eea

The momentum integration can be carry out by means of Schwinger's
known algebraic procedure \cite{Schwinger:1973kp} and converted
into an integration over two scalar para\-meters, $s$ and $t$. The
basic exponential is \be\Theta=e^{-sp^2}e^{-t(p-k)^2}\ee
and the  integration over the momentum $p$ is denoted by the average $\langle\dots\rangle$.
 The following well known formulas hold:
\bea \label{Theta}\langle\Theta\rangle=
\frac{\exp \left[-k\left(\frac{st}{s+t}\delta^{||}+\frac{ST}{S+T}\delta^{\perp}\right) k \right]}{(4\pi)^2(s+t)\sinh(s+t)}
\eea
with $S=\tanh(s)$ and $T=\tanh(t)$ and
\be\label{av1}   \langle p_\mu \ \Theta\rangle  =
\left(\frac{A}{D}k\right)_\mu \langle \Theta\rangle , \ee
\be\label{av2}   \langle p_\mu p_\nu \ \Theta\rangle  =
\left(\left(\frac{A}{D}k\right)_\mu \left(\frac{A}{D}k\right)_\nu
-i \left(\frac{F}{D^\tra}\right)_{\mu\nu}\right)\langle \Theta\rangle .\ee
The notations $A\equiv E^t-1$ and $D\equiv E^{s+t}-1 $ are used.
Explicit formulas are
\be\label{A/D}  \frac{A}{D}=\delta^{||}\frac{t}{s+t}
-iF\frac{\sinh(s)\sinh(t)}{\sinh(s+t)}
+\delta^\perp\frac{\cosh(s)\sinh(t)}{\sinh(s+t)} \ee
along with
\be\label{FED}\frac{-2iFE^{-s}}{D^{\top}}=\frac{\delta_{||}}{s+t}
-i F \frac{\sm}{\sp}+\delta_\perp \frac{\cm}{\sp}, \ee
where we dropped the indices. It should be remarked that all these matrices, i.e., $E$, $F$, $D$, commute.
In addition we need the relation
$$ p(s)_\mu\equiv e^{-sp^2}p_\mu e^{sp^2}={E^s}_{\mu\nu} p_\nu $$
for commuting a factor $p_\mu$ with the propagator $G(p)$.

Within this
formalism, the polarization tensor becomes an expression of the
type \be \label{Pist}\Pi_{\la\la'}(k)=\int_0^\infty\int_0^\infty \
ds \ dt \ \langle M_{\la\la'}(p,k) \Theta \rangle\ee
where in $M_{\la\la'}(p,k)$ we collected all factors appearing from the vertexes and from the lines except for that which go into $\Theta$. By using \Ref{av1} and \Ref{av2} the average, i.e., the momentum integration over $p$, can be transformed into
\be
\langle M_{\la\la'}(p,k) \Theta \rangle = M_{\la\la'}(s,t) \langle
\Theta \rangle ,\ee
where now $M_{\la\la'}(s,t)$ collects all factors except for $\langle\Theta\rangle $.

To make these transformations manageable we break the whole
polarization tensor into parts according to the division
introduced in Eq. \Ref{vertexfactor}. We get
\be\label{subdivision}
\Pi_{\la\la'}(k)=\sum_{i,j}\Pi^{ij}_{\la\la'}(k)+\Pi^{\rm ghost
}_{\la\la'}(k)\ee
with
\be\label{}\Pi^{ij}_{\la\la'}(k)=\Gamma^{(i)}_{\mu\nu\la}G_{\mu\mu'}(p)
\Gamma^{(j)}_{\mu'\nu'\la}G_{\nu'\nu}(p-k)
\ee
and corresponding subdivisions of $M$.

Now we calculate the individual contributions. The first is
\bea\label{Pi11}\Pi_{\la\la'}^{11}&=&
(k-2p)_\la G_{\mu\mu'}(p)(k-2p)_{\la'}G_{\mu'\mu}(p-k)\eea
and it transforms into
\bea\label{M11}
&& M_{\la\la'}^{11}(s,t)\langle\Theta\rangle =\left\langle (k-2p)_\la (k-2p(s))_{\la'}E^{-s}_{\mu\mu'}E^{-t}_{\mu'\mu} \ \Theta \right\rangle \nn\\
&&= \left[ \left(\left(1-2\frac{A}{D}\right)k\right)_{\la}\left(\left(1-2E^{s}\frac{A}{D}\right)
k\right)_{\la'} \right. \\ && ~~~~~~~~~~~~~~~~~~~~~~~~~~ \left.
-4i \left( \frac{E^{-s}F}{D^\top} \right)_{\la\la'}
\nn
\right]
tr E^{-s-t}
\left\langle \Theta \right\rangle .
\eea
We note the property $E^{s}\frac{A}{D}=\left(\frac{A}{D}\right)^\top$. The trace is
\bea\label{expf} tr E^{-s-t}&=& tr \left( \delta^{||}+iF\sinh(2(s+t))+\delta^{\perp} \cosh(2(s+t))\right)\nn\\
&=&2\left(1+\cosh(2(s+t))\right).
\eea

The second part in this expression can be integrated by parts. We represent
\bea\label{ds-dt}
\frac{-2iFE^{-s}}{D^{\top}}&=&\frac12\left(\frac{\pa}{\pa
s}-\frac{\pa}{\pa t}\right)   \left[
\frac{s-t}{s+t}\delta^{||}-iF\frac{\cm}{\sp} \right. \nn \\ &&
~~~~~~~~~~~~~~~~~~\left.+\delta^{\perp}\frac{\sm}{\sp}\right].
\eea The derivatives with respect to $s$ and $t$ will be
integrated by parts. We should note that expressions which are
symmetric under an exchange of $s$ and  $t$ are not affected by
the differentiation in Eq.\Ref{ds-dt}. So we have only the
derivative of the exponential in $\Theta$, \be
\label{deriv}\frac12\left(\pa_s-\pa_t\right)\left\langle \Theta
\right\rangle=\frac{1}{2} \ B_1 \  \left\langle \Theta
\right\rangle \ee with the notation
\be \label{B1a}B_1\equiv \frac{s-t}{s+t}l^2+\frac{\sm}{\sp}h^2 \ee
which will frequently appear in the following.

Using the notation of Eq.\Ref{ts1} we define
\bea\label{P11}P_\la&=&\left(\left(1-2\frac{A}{D}\right)k\right)_{\la}\nn\\
&=&\frac{s-t}{s+t} l_\la+
2i d_\la \frac{\ss\st}{\sp}+
h_\la \frac{\sm}{\sp} \nn\\
&\equiv&r  l_\la+i\al d_\la+\beta h_\la.
\eea
Using \Ref{ds-dt} and \Ref{deriv} for the integration by parts and
\Ref{P11} we represent
\bea\label{M11a} M^{11}_{\la\la'}(s,t)&=&\left\{P_\la
P^\top_{\la'}- \left[ \frac{s-t}{s+t}\delta_{||}-iF\frac{\cm}{\sp}
\right.\right.  \nn \\ &&   \left.\left.
+\delta_{\perp}\frac{\sm}{\sp}\right]_{\la\la'} B_1\right\}\nn \\
&& \times 2\left(1+\cosh(2(s+t))\right)\eea The contribution with
the field strength $F$ disappears if taking into account  the
symmetry of the integrals over $s$ and $t$ under exchange of these
two arguments. Applying formula \Ref{rep1} with
$a=\frac{s-t}{s+t}B_1$ and $b=\frac{\sm}{\sp}B_1$ we obtain
\bea\label{M11b} &&M^{11}(s,t)   \\ &&=\left\{
-\left(\frac{s-t}{s+t}\right)^2T^{(1)} \right.  \nn \\ && \left.
+\left[
\left(\frac{2\ss\st}{\sp}\right)^2-\left(\frac{\sm}{\sp}\right)^2\right]T^{(2)}\right.
\nn \\ && \left.- \frac{s-t}{s+t} \ \frac{\sinh(s-t)}{\sp} T^{(3)}
\right\} 2\left(1+\cosh(2(s+t))\right).\nn\eea
All other contributions canceled.

Next contribution is $M^{22}$. From \Ref{vertexfactor} we have
\bea\label{Pi22}
\Pi^{22}_{\la\la'}&=&4\left(g_{\la\mu}k_{\nu}-g_{\la\nu}k_{\mu}\right)G_{\mu\mu'}(p)
\nn \\ && ~~~~\left(g_{\la'\mu'}k_{\nu'}-g_{\la'\nu'}k_{\mu'}\right)G_{\nu'\nu}(p-k)
\eea
and
\bea\label{M22}M^{22}_{\la\la'}(s,t)&=&4\left(
E^{-s}_{\la\la'}(kE^{-t}k)+E^t_{\la\la'}(kE^{-s}k)
\right. \\ && \left.-\left(E^{-s}k\right)_{\la}\left(E^{-t}k\right)_{\la'}
-\left(E^{t}k\right)_{\la}\left(E^{s}k\right)_{\la'}\right).\nn
\eea
Integration by parts is not required here. We introduced the short notation
\bea\label{kEk}(kE^{-t}k)\equiv k_\la E^{-t}_{\la\la'}k_{\la'}
= l^2+h^2\cosh(2t). \eea
Here we have an expression of the type of \Ref{ts2}. The parameter are
\be \begin{array}{rclrclrcl}
r&=&1,&\al&=&\sinh(2t),&\beta&=&\cosh(2t)\\
s&=&1,&\gamma&=&\sinh(2s),&\delta&=&\cosh(2s)
\end{array}
\ee
and
\beao a&=&-2l^2-h^2(\cosh(2s)+\cosh(2t))\\
b&=&-l^2(\cosh(2s)+\cosh(2t))-2h^2\cosh(2s)\cosh(2t).\eeao The
tranversality condition is fulfilled and from \Ref{rep2} we obtain
\bea\label{M22b} M^{22}(s,t)&=&8T^{(1)}+8\cosh(2(s+t))T^{(2)}
\nn \\ &&
+4(\cosh(2s)+\cosh(2t))T^{(3)} . \eea

Next we consider $\Pi^{12}$ and $\Pi^{21}$. From \Ref{vertexfactor} we get
\bea\label{Pi12}
&&\Pi^{12}_{\la\la'}+\Pi^{21}_{\la\la'}   =\\&&
 g_{\mu\nu}(k-2p)_{\la}
G_{\mu\mu'}(p)
2\left(g_{\la'\mu'}k_{\nu'}-g_{\la'\nu'}k_{\mu'}\right)G_{\nu'\nu}(p-k)\nn\\&&+
2\left(g_{\la\mu}k_\nu-g_{\la\nu}k_\mu\right)
G_{\mu\mu'}(p)
 g_{\mu'\nu'}(k-2p)_{\la'}G_{\nu'\nu}(p-k),\nn
\eea
and further
\bea\label{M12}&&
\left(M^{12}_{\la\la'}+M^{21}_{\la\la'} \right)\langle\Theta\rangle   \nn =\\&&
\left\langle \left\{ 2(k-2p)_{\la}E^{-s}_{\nu\la'}k_{\nu'}E^{-t}_{\nu'\nu}
-2(k-2p)_{\la}E^{-s}_{\nu\mu'}k_{\mu'}E^{-t}_{\la'\nu}  \right. \right. \nn \\ &&\left.\left.
+2k_{\nu}E^{-s}_{\la\nu'}(k-2p(s))_{\la'}E^{-t}_{\nu'\nu}
 \right. \right. \nn \\ &&\left.\left.-2k_{\mu}E^{-s}_{\mu\nu'}(k-2p(s))_{\la'}E^{-t}_{\nu'\la}  \right\}
 \Theta\right\rangle \nn \\ &=& \left\langle
2\left\{  (k-2p)_{\la} \left(\left(E^{s+t}-E^{-s-t}\right)k\right)_{\la'} \nn \right.\right. \\&&\left. \left.
        + \left(\left(E^{s+t}-E^{-s-t}\right)^\top k\right)_{\la} (k-2p(s))_{\la'} \right\}
        \Theta\right\rangle .
  \eea
We use the averages \Ref{av1} and obtain an expression of the form of \Ref{ts2}
with
\bea\label{}&&P=1-2\frac{A}{D} \nn \\ && =\frac{s-t}{s+t}\delta^{||}+iF\frac{2\ss\st}{\sp}+
\delta^{\perp}\frac{\sm}{\sp} ,\nn
\\&&Q=E^{s+t}-E^{-s-t}=2iF\sinh(2(s+t))
\eea and from \Ref{rep2} we find simply
\be\label{M12b}
M^{12}(s,t)+M^{21}(s,t) =-32 \cp \ss\st \ T^{(2)}  .\ee

Next we consider the contribution of $\Pi^{33}$ together with the
contribution  from the ghosts, $\Pi^{\rm ghost}$. We get
\bea\label{Pi33}\Pi^{33}&=& \left(
g_{\la\mu}(p-k)_\nu+g_{\la\nu}p_{\mu}\right)G_{\mu\mu'}(p)
 \nn \\ && \left(g_{\la'\mu'}(p-k)_{\nu'}+g_{\la'\nu'}p_{\mu'}\right)G_{\nu'\nu}(p-k).
\eea We use the property of the propagator
\be\label{cp} p_{\mu}G_{\mu\mu'}(p)=G(p)p_{\mu'} \ee
and obtain after simple calculation, using for instance the cyclic property of the trace,
\bea\label{Pi33a}\Pi^{33}_{\la\la'}&=&G_{\la\la'}(p)G(p-k)(p-k)^2\nn
\\ &&+p^2G(p)G_{\la'\la}(p-k) \nn\\ &&
+p_{\la}G(p)(p-k)_{\la'}G(p-k)\nn \\
&&+(p-k)_{\la}G(p)p_{\la'}G(p-k).\nn \eea In the first two lines
in the r.h.s. one line collapses into a point by means of, e.g.,
$p^2G(p)=1$ and the corresponding graph becomes a tadpole which we
do not consider here. The second line in the r.h.s. is just equal
to the contribution from the ghosts, second line in \Ref{NPT1},
with opposite sign and cancels. So we obtain simply
\be\label{Pi33+ghosts}\Pi^{33}+\Pi^{\rm ghost}=0.\ee

Next we consider $\Pi^{13}$ and $\Pi^{31}$. We start from
\bea\label{Pi13} \Pi^{13}_{\la\la'}&=&
g_{\mu\nu}(k-2p)_{\la}G_{\mu\mu'}(p)\nn \\ &&\left(g_{\la'\mu'}(p-k)_{\nu'}+
g_{\la'\nu'}p_{\mu'}\right)G_{\nu'\nu}(p-k) , \nn
\\
\Pi^{31}_{\la\la'}&=&
\left(g_{\la\mu}(p-k)_{\nu}+g_{\la\nu}p_{\mu}\right)\nn \\ &&G_{\mu\mu'}(p)
g_{\mu'\nu'}(k-2p)_{\la'}G_{\nu'\nu}(p-k)
\eea
and arrive at
\beao\label{M13}
M^{13}_{\la\la'}(s,t)&=&-\left\langle(k-2p)_{\la} \left(E^{\rm sy}(k-2p(s))+E^{\rm as}k\right)_{\la'}
\Theta \right\rangle ,\nn \\
M^{31}_{\la\la'}(s,t)&=&-\left\langle \left(E^{\rm sy}(k-2p)-E^{\rm as}k\right)_{\la}(k-2p(s))_{\la'}
\Theta \right\rangle,
\eeao
where the notations
\bea\label{Eas}
E^{\rm sy}&=&\frac12\left(E^{s+t}+E^{-s-t}\right)=\delta^{||}+\delta^{\perp}\cosh(2(s+t))\nn \\
E^{\rm as}&=&\frac12\left(E^{s+t}-E^{-s-t}\right)=iF\sinh(2(s+t))
\eea
have been introduced.

The averages can be calculated and we obtain
\bea\label{M13a}
M^{13}_{\la\la'}(s,t)&=&\left\{-P_{\la}Q_{\la'}+4i\left(\frac{E^{\rm sy}E^{-s}F}{D^\top}\right)_{\la\la'}\right\} \langle\Theta \rangle\nn ,\\
M^{31}_{\la\la'}(s,t)&=&\left\{-Q^\top_{\la}P^\top_{\la'}+4i\left(\frac{E^{\rm sy}E^{-s}F}{D^\top}\right)_{\la\la'}\right\} \langle\Theta \rangle \nn \\
\eea
with $P_\la$ given by Eq.\Ref{P11} and
\bea\label{88}
Q_{\la'}&=&\left(\left(E^{\rm as}+E^{\rm
sy}\left(1-2\frac{A}{D}\right)^\top \right) k\right)_{\la'}\nn \\
&=& \frac{s-t}{s+t}l_{\la'}+id_{\la'}\Bigg(\sinh(2(s+t)) \nn \\ &&
-\frac{2\ss\st\cosh(2(s+t))}{\sp}\Bigg)\nn \\ &&
+h_{\la'}\frac{\sm}{\sp}\cosh(2(s+t)) \nn\\&\equiv& s\
l_\la+\gamma \ id_\la+ \delta \ \ h_\la. \eea The second
contributions to the r.h.s. in Eq.\Ref{M13a} must be integrated by
parts.  Using \Ref{ds-dt} and \Ref{deriv} we get
\bea\label{M13b}&&
M^{13}_{\la\la'}(s,t)+M^{31}_{\la\la'}(s,t) =\Bigg\{
-P_{\la}Q_{\la'}-Q^\top_{\la}P^\top_{\la'} \\ &&+
2\left[  \frac{s-t}{s+t}\delta^{||}_{\la\la'}+
\frac{\sm}{\sp}\cosh(2(s+t))\delta^\perp_{\la\la'}
\right] B_1 \Bigg\}  .\nn
\eea
The contribution with $F_{\la\la'} $ has been dropped again for
symmetry reasons. Now we apply \Ref{ts2}  with $r,\al,\beta$ given
by Eq.\Ref{P11},   with $s,\gamma,\delta$ given by Eq.\Ref{88} and
$a$, $b$ following from Eq.\Ref{M13b},
\bea\label{} a&=&-2\frac{s-t}{s+t}\left[ \frac{s-t}{s+t} l^2+\frac{\sm}{\sp}h^2\right] \nn \\
b&=&-2\frac{\sm}{\sp}\cosh(2(s+t)) \left[ \frac{s-t}{s+t} l^2
\right. \nn\\ && \left. ~~~~~~~~~~~~~~~~~~
+\frac{\sm}{\sp}h^2\right],
\eea
and obtain
\bea\label{Pi13a}&&
\Pi^{13}+\Pi^{31}=2\left(\frac{s-t}{s+t}\right)^2T^{(1)} \nn \\&&
+
2\left\{\left[\left(\frac{\sm}{\sp}\right)^2-
\left(\frac{2\ss\st}{\sp}\right)^2\right]\right. \nn \\ && \left.~~~~~~~~~~~~~~~~~~~~~~~~~~~~\times \cosh(2(s+t))\right. \nn \\ && \left. +
\frac{2\ss\st}{\sp} \sinh(2(s+t))\right\}T^{(2)} \nn \\
&&+\frac{s-t}{s+t} \ \frac{\sm}{\sp}\left(1+\cosh(2(s+t))\right) \
T^{(3)}
\nn \\ &&
-2\frac{s-t}{s+t}\ \sp\sm T^{(5)}
.
\eea

Finally we need $\Pi^{23}$ and $\Pi^{32}$. Proceeding in the same
way as before we derive from \Ref{NPT1} and \Ref{vertexfactor}
\bea\label{Pi23}
\Pi^{23}_{\la\la'}&=&2(g_{\la\mu}k_{\nu}-g_{\la\nu}k_{\mu})G_{\mu\mu'}(p)
\nn \\ &&
~~~~~~~(g_{\la'\mu'}(p-k)_{\nu'}+g_{\la'\nu'}p_{\mu'})G_{\nu'\nu}(p-k) \nn \\
\Pi^{32}_{\la\la'}&=&2(g_{\la\mu}(p-k)_{\nu}+g_{\la\nu}p_{\mu})G_{\mu\mu'}(p)
\nn \\ &&
~~~~~~~(g_{\la'\mu'}k_{\nu'}-g_{\la'\nu'}k_{\mu'})
G_{\nu'\nu}(p-k)
\eea
which gives
\bea\label{M23a}&&
M^{23}_{\la\la'}(s,t)\langle\Theta\rangle
=\nn \\ &&
\bigg\langle 2\bigg\{
-E^{-s}_{\la\la'}\left(kE^{t}(k-p(s))\right)
-E^{t}_{\la\la'}\left(kE^{-s}p(s)\right) \nn \\ &&
+\left(E^{-s}p(s)\right)_{\la}\left(E^{-t}k\right)_{\la'}
+\left(E^{t}(k-p(s))\right)_{\la}\left(E^{s}k\right)_{\la'}
\bigg\} \Theta\bigg  \rangle , \nn \\
&&M^{32}_{\la\la'}(s,t)\langle\Theta\rangle
=\nn \\ &&
\bigg\langle 2\bigg\{
-E^{-s}_{\la\la'}\left(kE^{-t}(k-p)\right)
-E^{t}_{\la\la'}\left(pE^{-s}k)\right) \nn \\ &&
+\left(E^{-s}k\right)_{\la}\left(E^{-t}(k-p)\right)_{\la'}
+\left(E^{t}k\right)_{\la}\left(E^{s}p\right)_{\la'}
\bigg\} \Theta\bigg\rangle . \eea
Now the average \Ref{av1} is used and the expression rearranged according to
\be\label{rearr}M^{23}_{\la\la'}+M^{32}_{\la\la'}=2(A+B)\ee
with
\bea\label{A}
A&=&-2E^{t}_{\la\la'}\left(k\frac{E^t-1}{D}k\right)
+\left(E^{t}k\right)_{\la}\left(\left(\frac{E^{t}-1}{D}\right)^\top k\right)_{\la'}
 \nn \\ &&
 +\left( \frac{E^{t}-1}{D}  k\right)_{\la}\left(E^{-t}k\right)_{\la'} , \nn \\
B&=&-2E^{-s}_{\la\la'}\left(k\frac{E^s-1}{D}k\right)
+\left(E^{-s}k\right)_{\la}\left(\frac{E^{s}-1}{D} k\right)_{\la'}
 \nn \\ &&
+\left( \left(\frac{E^{s}-1}{D} \right)^\top  k\right)_{\la}\left(E^{s}k\right)_{\la'}   . \nn
\eea
$A$ and $B$ have the structure of \Ref{ts2}. For $A$ we get
\be\label{A1}
\begin{array}{rclrclrcl}
r&=&1, &\al&=&\sinh(2t),&\beta&=&\cosh(2t) \\
s&=&\frac{t}{s+t},&\gamma&=&\frac{\ss\st}{\sp},&\delta&=&\frac{\cs\st}{\sp}\end{array}
\ee
and
\bea
a&=&-2\left(\frac{t}{s+t}l^2+\frac{\cs\st}{\sp}h^2\right), \nn \\
b&=&-2\cosh(2t)\left(\frac{t}{s+t}l^2+\frac{\cs\st}{\sp}h^2\right) \nn \\
c&=&-2\sinh(2t)\left(\frac{t}{s+t}l^2+\frac{\cs\st}{\sp}h^2\right)
\eea
and for $B$
\be\label{B1}
\begin{array}{rclrclrcl}
r&=&1, &\al&=&-\sinh(2s),&\beta&=&\cosh(2s) \\
s&=&\frac{s}{s+t},&\gamma&=&-\frac{\ss\st}{\sp},&\delta&=&\frac{\ss\ct}{\sp}\end{array}
\ee
and
\bea
a&=&-2\left(\frac{s}{s+t}l^2+\frac{\ss\ct}{\sp}h^2\right), \nn \\
b&=&-2\cosh(2s)\left(\frac{s}{s+t}l^2+\frac{\ss\ct}{\sp}h^2\right) ,\nn \\
c&=&2\sinh(2s)\left(\frac{s}{s+t}l^2+\frac{\ss\ct}{\sp}h^2\right).
\eea
Putting these contributions together we obtain with \Ref{rep2}
\bea\label{Pi23a}&&
\Pi^{23}+\Pi^{32}=2\Bigg\{-2T^{(1)} \nn \\
&&-2\Bigg[\frac{\cosh(2t)\cs\st+\cosh(2s)\ss\ct}{\sp} \nn \\ && +2\cosh(s-t)\ss\st\Bigg]
T^{(2)} \nn \\ &&
-\left[1+\frac{s\cosh(2s)+t\cosh(2t)}{s+t}\right]
T^{(3)} \nn \\ &&
+\left[-1+\frac{s\cosh(2s)+t\cosh(2t)}{s+t}\right]
T^{(5)}   \Bigg\} .
\eea

In this way the contributions of the individual parts are
calculated. Putting all together we get from \Ref{M11b},
\Ref{M12b}, \Ref{Pi13a} and \Ref{Pi23a} the final expression which
we represent in terms of form factors according to \Ref{exp},
\be\label{formfM} \Pi^{(i)}(k)=\int_0^\infty\int_0^\infty \ ds \ dt \ M^{(i)}(s,t) \ \langle\Theta\rangle
\ee
with
\bea\label{Pifinal}
&&M^{(1)}(s,t)=4-2\left(\frac{s-t}{s+t}\right)^2 \cosh(2(s+t)), \nn \\
&&M^{(2)}(s,t)=
\\ &&
2\left[\left(\frac{2\ss\st}{\sp}\right)^2-
\left(\frac{\sm}{\sp}\right)^2\right]  \nn \\ &&
-4+8\cosh(s-t)\cosh(s+t)
 ,\nn \\
&&M^{(3)}(s,t)= \nn \\ &&  -2\frac{s-t}{s+t}\left[\frac{\sm}{\sp}\frac{1+\cosh(2(s+t))}{2}+
          \right.\nn\ \\ && \left.~~~~~~~~~~~~~~~~~~~~~~~~~~~~~~~~~  \frac{\cosh(2s)-\cosh(2t)}{2}\right]
\nn \\ &&
            -2+3(\cosh(2s)+\cosh(2t)) ,\nn \\
&&M^{(5)}(s,t)=-2\left(1-\frac{\cosh(2s)+\cosh(2t)}{2}\right),
\eea where we made again use of the symmetry under
$s\leftrightarrow t$ dropping the  contribution proportional to
$T^{(3)}$ in Eq.\Ref{M22b}.

For vanishing magnetic field $B\to 0$ which means to take the
lowest order in $s$ and $t$ we obtain \bea\label{zeroth}&&
\Pi_{\la\la'}(k)=\int_0^\infty\int_0^\infty \ ds \ dt \
2\left(2-\left(\frac{s-t}{s+t}\right)^2\right) \nn \\ &&
~~~~~~~~~~~~~~~~~~\times\left(T^{(1)}_{\la\la'}+T^{(2)}_{\la\la'}+
T^{(3)}_{\la\la'}\right) \langle\Theta\rangle,\eea where here we
have simply  $\langle\Theta\rangle=(4\pi(s+t))^{-2} \
\exp(-\frac{st}{s+t}k^2)$ instead of \Ref{Theta}.

\subsection{Neutral Polarization Tensor in Fujikawa Gauge}\label{sectFJ}
With the derived in the preceding section formulas at hand it is
easy to calculate the neutral polarization tensor in Fujikawa
gauge (for a definition see \cite{}). In that gauge the third part of the vertex factor,
$\Gamma^{(3)}_{\mu\nu\la}$, is absent and the ghost contribution
looks different,
\be\label{ghostinFujikawa}
\Pi^{\rm ghost FJ}_{\la\la'}=(k-2p)_\la G(p)(k-2p)_{\la'}G(p-k).\ee
It differs from the contribution $\Pi^{11}_{\la\la'}$, Eq. \Ref{Pi11}, by the different lines only. While in \Ref{Pi11} the lines belong to a vector particle, in \Ref{ghostinFujikawa} they belong to a scalar one. The vertex factors in \Ref{Pi11} and in \Ref{ghostinFujikawa} are the same. So the whole difference is in the absence of the factor given by Eq. \Ref{expf} and in the different sign   the ghosts enter with. In this way we get
\bea\label{MghostFj}&&
M^{\rm ghost FJ}=-2\left\{
-\left(\frac{s-t}{s+t}\right)^2T^{(1)}
\right. \\ && \left.+\left[ \left(\frac{2\ss\st}{\sp}\right)^2-\left(\frac{\sm}{\sp}\right)^2\right]T^{(2)}\right.
\nn \\ && \left.-
\frac{s-t}{s+t}\frac{2\ss\st}{\sp} T^{(3)}  \right\} .\eea
The polarization tensor in Fujikawa gauge is
\be\label{} \Pi^{\rm Fj gauge}_{\la\la'}=
\Pi^{11}_{\la\la'}+\Pi^{22}_{\la\la'}+\Pi^{12}_{\la\la'}+\Pi^{21}_{\la\la'}+\Pi^{\rm ghost FJ}_{\la\la'}\ee
Taking together \Ref{M11b} and \Ref{MghostFj} along with
\Ref{M22b} and \Ref{M12b} and using a representation like
\Ref{formfM} and \Ref{exp} we obtain for the form factors
\bea\label{PifinalFj}
M_{\rm Fj}^{(1)}&=&  8 -2\left(\frac{s-t}{s+t}\right)^2   \cosh(2(s+t))   ,     \\
M_{\rm Fj}^{(2)}&=& 2\left(4+
\left(\frac{2\ss\st}{\sp}\right)^2   -\left(\frac{\sm}{\sp}\right)^2 \right)
\nn \\ && ~~~~~~~~~~~~~~~~~~~~\times\cosh(2(s+t))   \nn \\ &&  -32\cosh(s+t)\ss\st  +8\cosh(2(s+t)) ,  \nn \\
M_{\rm Fj}^{(3)}&=&   -2\frac{s-t}{s+t} \
\frac{\sm}{\sp}\cosh(s(s+t))\nn \\ && ~~~~~~~~~~~~~+4\left(\cosh(2s)+\cosh(2t)\right)
.      \nn \eea In the limit without magnetic field we get in this
case
\bea\label{zerothFj}
&&\Pi^{\rm Fj}_{\la\la'}(k)=\int_0^\infty\int_0^\infty \ ds \ dt \ 2\left(4-\left(\frac{s-t}{s+t}\right)^2\right)\nn \\ && ~~~~~~~~~~~~~~~~\times \left(T^{(1)}_{\la\la'}+T^{(2)}_{\la\la'}+
T^{(3)}_{\la\la'}\right)\langle\Theta\rangle,\eea
a result which is different from \Ref{zeroth}.


\subsection{The Non-Transversality of the Neutral Polarization Tensor}\label{sectnontr}
The non-transversality of the neutral polarization tensor can be established by direct calculation before performing the momentum integration. For this to do we start from Eq. \Ref{NPT1} and consider $\Pi_{\la\la'}(k) \ k_{\la'}$. We use the property of the vertex factor
\be\label{Gak1}\Gamma_{\mu'\nu'\la'}k_{\la'}=K_{\mu'\nu'}(p-k)-K_{\mu'\nu'}(p)
\ee
with $K_{\mu\nu}(p)$ defined by \Ref{K}. We get
\bea\label{Pik1}&&\Pi_{\la\la'}(k) \ k_{\la'}=\\&&
\Gamma_{\mu\nu\la}G_{\mu\mu'}(p)\left[ K_{\mu'\nu'}(p-k)-K_{\mu'\nu'}(p) \right]G_{\nu\nu'}(p-k) \nn \\ &&
-p_\mu\frac{1}{p^2}\ (p-k)k \ \frac{1}{(p-k)^2}
-(p-k)_\mu\frac{1}{p^2}\ pk \  \frac{1}{(p-k)^2}.\nn
\eea
We proceed with two times using the obvious relation (see Eq.\Ref{prop})
\be\label{obr}K_{\mu\la}(p) \ G_{\la\nu}(p)=\delta_{\mu\nu}-p_\mu \ \frac{1}{p^2} \ p_\nu .
\ee
There will be two contributions resulting from $\delta_{\mu\nu}$ in \Ref{obr}. In these one denominator disappears and  we drop these contributions. This is for the same reason as discussed in the end of section 2. In the ghost terms we use $(p-k)p=\frac12(p^2-k^2-(p-k)^2)$ and $pk=\frac12(p^2+k^2-(p-k)^2)$. Again, the terms where one denominator cancels are dropped and we have to keep only what is proportional to $k^2$. We obtain
\bea\label{Pik2}&&\Pi_{\la\la'}(k) \ k_{\la'}=
-\Gamma_{\mu\nu\la}G_{\mu\mu'}(p)(p-k)_{\mu'}\frac{1}{(p-k)^2}(p-k)_{\nu} \nn \\ &&
+ \Gamma_{\mu\nu\la}p_{\mu}\frac{1}{p^2}p_{\nu'}G_{\nu'\nu}(p-k)+\frac12k_\la k^2\frac{1}{p^2(p-k)^2}.
\eea
Next we use the cyclic property of the trace and move in the first
term the factor $(p-k)_\nu$ to the left side. After that we use
$\Gamma_{\mu\nu\la}p_\mu=K_{\nu\la}(p-k)-K_{\nu\la}(k)$
 and get
\bea\label{Pik3}&&\Pi_{\la\la'}(k) \ k_{\la'}=\\&&
\left[ K_{\la\mu}(p)-K_{\la\mu}(k) \right]G_{\mu\mu'}(p) (p-k)_{\mu'}\frac{1}{(p-k)^2} \nn \\ &&
-\left[ K_{\nu\la}(p-k)-K_{\nu\la}(k) \right]\frac{1}{p^2}p_{\nu'}G_{\nu'\nu}(p-k)  \nn \\ && +\frac12k_\la k^2\frac{1}{p^2(p-k)^2}.
\eea
We claim this expression is equal to
\bea\label{Pik4}\Pi_{\la\la'}(k) \ k_{\la'}&=&
K_{\la\mu}(k) \ \Sigma_\mu
\eea
with
\bea\label{Sigmu}&&\Sigma_\mu\equiv\\&&\left[ G_{\mu\mu'}(p)(p-k)_{\mu'}\frac{1}{(p-k)^2}-\frac{1}{p^2}p_{\nu'}G_{\nu'\mu}(p-k)   \right].\nn
\eea
To obtain this we used for the first and for the third terms in
\Ref{Pik3} the property \Ref{obr} and after that
$p(p-k)=\frac12(p^2-k^2+(p-k)^2)$ from which we keep $k^2$ only.
Then it is seen  that what remains compensates the contribution
resulting from the ghosts (last term in r.h.s. on \Ref{Pik3}).

Now it is obvious that $k_\la\Pi_{\la\la'}(k) \ k_{\la'}=0$ holds.
The r.h.s. of \Ref{Sigmu} can be simplified. We use
\be\label{pGGp}p_\mu G_{\mu\mu'}(p)=\frac{1}{p^2}p_{\mu'}
\ee
which is essentially a special case of \Ref{cp} and get
\bea&&\label{Sigmu1}\Sigma_\mu=\\&&k_\mu \frac{1}{p^2(p-k)^2}-G_{\mu\mu'}(p)k_{\mu'}\frac{1}{(p-k)^2}-
\frac{1}{p^2}k_{\nu'}G_{\nu'\mu}(p-k).\nn
\eea
Now we turn to the representation with the parametric integrals introduced in section \ref{sect2}, especially the formulas \Ref{prop0} and \Ref{prop}. We obtain
\bea\label{}\Sigma_\mu&=&\int ds\int dt \ \left(k_\mu-E^{-s}_{\mu\mu'}k_{\mu'}-k_{\nu'}E^{-t}_{\nu'\la}\right)\langle\Theta\rangle
\nn \\ &=& \left(\left(-E^{-s}-E^t+1\right)k\right)_\mu \langle\Theta\rangle
\eea
with $\langle\Theta\rangle$ given by Eq.\Ref{Theta}.

In order to multiply with $K_{\la\mu}(k)$ we note
\be\label{}K_{\la\mu}(k)l_\mu=T^{(5)}_{\la\la'}k_{\la'} \mbox{ and }
K_{\la\mu}(k)h_\mu=-T^{(5)}_{\la\la'}k_{\la'}
\ee
In this way we get finally
\bea\label{}&&\Pi_{\la\la'}(k) \ k_{\la'}=\\&&\int ds\int dt \ (-2+\cs+\ct)\langle\Theta\rangle
 \ T^{(5)}_{\la\la'}k_{\la'}.\nn
\eea
This corresponds just to the contribution of $M^{(5)}(s,t)$ in \Ref{Pifinal}.

\section{The Charged Polarization Tensor}
The charged polarization tensor is shown in Fig. \ref{figure:Picha}. It is denoted by $\Pi_{\la\la'}(p)$ where the argument $p$ is a momentum defined in \Ref{defp}. The calculations in this section are to a large extend in parallel to that in the preceding one for the neutral polarization tensor, however different in the details. For a number of objects we use here the same notations as in that section, however they have a different meaning and they are valid in this section only.
The magnetic field is  equal to unity in this section too.
\subsection{Operator Structures}
The general structure of the charged polarization tensor is defined by its property not to be transversal but to obey the weaker condition
\be\label{doubletransvp} p_\la \ \Pi_{\la\la'}(p) \ p_{\la'}=0 . \ee
In this section we introduce the vectors $l_\mu$ and $h_\mu$ by
\bea&&\label{vecp} l_\mu=\left(\begin{array}{c}0\\0\\p_3\\p_4\end{array}\right),~
h_\mu=\left(\begin{array}{c}p_1\\p_2\\0\\0\end{array}\right), ~ d_\mu=\left(\begin{array}{c}p_2\\-p_1\\0\\0\end{array}\right),~ \nn\\&&F_{\mu\lambda}=\left(\begin{array}{cccc}0&1&0&0\\-1&0&0&0\\0&0&0&0\\0&0&0&0\end{array}
\right),\eea
where the third vector is  $d_\mu\equiv F_{\mu\nu}p_\nu$. These notations are valid in the present section only. Actually they are different from that in the preceding section  by the noncommutativity of $p$, Eq.\Ref{noncp}, only. We note $p_\la=l_\la+h_\la$ and the noncommutative pice is $h_\mu$ for which $[h_\mu,h_\nu]=iF_{\mu\nu}$ holds. It must be remarked that the $h_\mu$ are operators. Later, when they act on the states like \Ref{chstates} the eigenvalues can be substituted according to $h^2\to 2n+1$. We make a distiction between the operator and its eigenvalues only where it is necessary for understanding.  Furthermore it is useful to collect  the following relations,
\be\label{dpcomms}
\begin{array}{rclrclrcl}
h_\la h_\la&=& h^2 , & d_\la d_\la&=&h^2 ,  &&& \\
{[}h_\la,h^2]&=&2id_\la ,& [d_\la,h^2]&=&-2ih_\la,\\
h_\la d_\la&=&i, & F_{\la\la'}d_{\la'}&=&-h_\la ,& F_{\la\la'}h_{\la'}&=&d_\la , \\
d_\la h_\la&=&-i ,& d_\la F_{\la\la'}&=&h_{\la'}, & h_\la F_{\la\la'}&=&-d_{\la'}. \end{array} \ee
Now the general structure allowed by the remaining Lorentz invariance is
\begin{eqnarray} \label{Tnp}
T^{(1)}_{\la\la'}&=& l^2 \delta^{||}_{\la\la'}-l_\la l_{\la'}\nn \\
T^{(2)}_{\la\la'}&=&h^2\delta^\perp_{\la\la'}+2iF_{\la\la'}
-h_\la h_{\la'}=d_\la d_{\la'} +iF_{\la\la'}   \nn \\
T^{(3)}_{\la\la'}&=& h^2 \delta^{||}_{\la\la'}+
l^2 \delta^\perp_{\la\la'} -l_\la h_{\la'}-h_\la l_{\la'}\nn \\
T^{(4)}_{\la\la'}&=& i(l_\la d_{\la'}-d_\la l_{\la'} )
+il^2F_{\la\la'}-\delta^{||}_{\la\la'}\nn \\
T^{(5)}_{\la\la'}&=&h^2 \delta^{||}_{\la\la'}-l^2 \delta^\perp_{\la\la'} \nn \\
T^{(6)}_{\la\la'}&=&
\delta^{||}_{\la\la'}+3\delta^{\perp}_{\la\la'}+(l^2+h^2)iF_{\la\la'}
\end{eqnarray}
and the identity $i(d_{\la}h_{\la'}-h_{\la}d_{\la'})=ih^2F_{\la\la'}+\delta^\perp_{\la\la'}$ holds.

As in the preceding section the first four operators are transversal, $p_\la T^{(i)}_{\la\la'}=T^{(i)}_{\la\la'}p_{\la'}=0$ for $i=1,2,3,4$ and
 the remaining two fulfill \Ref{doubletransvp} only, $p_\la T^{(i)}_{\la\la'}p_{\la'}=0$ for $i=5,6$. The sum of the first three operators is just the transversal part of the kernel of the quadratic part of the action, Eq. \Ref{Kp},
\be\label{sumTipa}
T^{(1)}_{\la\la'}+T^{(2)}_{\la\la'}+T^{(3)}_{\la\la'}=K_{\la\la'}(p).
\ee

The first three operators, $T^{(i)}_{\la\la'}$ with $i=1,2,3$, commute. Hence they have common eigenvectors. These are some linear combinations of just the tree level states $\mid p_{||},n,s\rangle_\alpha$, \Ref{chstates}, of the charged gluons defined in Sect. \ref{bn}.
The following relations hold,
\be \label{evs}\begin{array}{rclrcl}
T^{(1)} \mid 1\rangle&=&0  , &\\ [3pt]
T^{(2)}\mid 1 \rangle&=&h^2\mid 1 \rangle ,\\ [3pt]
T^{(3)}\mid 1 \rangle&=&l^2\mid 1 \rangle ,\\ [3pt]
K\mid 1 \rangle&=&(l^2+h^2)\mid 1 \rangle ,&\\ [3pt]
T^{(1)}\mid 2 \rangle&=&l_3l_4\frac{h}{k}\mid 0 \rangle+\frac{l_4^2h^2}{k^2}\mid  2\rangle-\frac{l_3l_4^2h}{k^2}\mid 3 \rangle \\ [3pt]
T^{(2)}\mid 2 \rangle&=&0,\\ [3pt]
T^{(3)}\mid 2 \rangle&=&-l_3l_4\frac{h}{k}\mid  0 \rangle+\left(\frac{l_3^2l_4^2}{k^2}+k^2\right)\mid 2 \rangle+\frac{l_3l_4^2}{k^2}\mid 3 \rangle \\ [3pt]
 K\mid 2 \rangle&=&(l^2+h^2)\mid 2 \rangle
\end{array} \ee
with $k=\sqrt{l_3^2+h^2}$ and, because acting on the states, the operator $h$ can be substituted by its eigenvalue, $h=\sqrt{2n+1}$.
We dropped the Lorentz indicees and some quantum numbers in the states. For example, the last term in the last line reads $K_{\al\beta}(p)\mid p_{||},n,2\rangle_\beta=(l^2+h^2)\mid p_{||},n,2\rangle_\alpha$
if all notations are kept.
The tensors $T^{(i)}_{\la\la'}$ rotated according to \Ref{Bma} in the plane perpendicular to the magnetic field are shown in the Appendix  where also the matrix elements of the remaining tensors are listed.

From the calculation of the graph in the next subsection contributions will appear which have the same structure as given by the formulas \Ref{Pla}, \Ref{ts2} and \Ref{Qla} in the preceding section and the coefficients in front of $T^{(1\dots 4)}$ in Eq. \Ref{rep1} are valid in this section too.

Before introducing the form factors a general remark is in order. The commutator relation \Ref{noncp} can  be rewritten in the form $p_\la p^2=(p^2\delta_{\la\la'}+2iF_{\la\la'})p_{\la'}$. As a consequence for any function of $p^2$ the relations
\be\label{p2+2iF}
p_\la f(p^2)=f(p^2+2iF)_{\la\la'}p_{\la'}
\ee
and
\be\label{p2+2iFa}
f(p^2)p_\la =p_{\la'}f(p^2+2iF)_{\la'\la}
\ee
hold where now $f$ must be viewed as a function of a matrix so that it itself becomes a matrix carrying the indices $\la$ and $\la'$. The same is true with $h^2$ in place of $p^2$,

Now the decomposition into form factors can be written as
\be\label{expp} \Pi_{\la\la'}(p)=\sum_{i=1}^6 \ \Pi^{(i)}(l^2,h^2+2iF)_{\la\la''} \  T^{(i)}_{\la''\la'} .\ee
The property \Ref{p2+2iF} ensures that \Ref{doubletransvp} holds.

\subsection{Strategy for the Calculation of the Charged Polarization Tensor}
As compared with the neutral polarization tensor  the calculation of the charged one is more difficult. A main complication is that it is unknown how to do the integration by parts which for the neutral tensor was done by the formulas \Ref{ds-dt} and \Ref{deriv}. Fortunately there is a way out. First of all the contributions from the graph can be divided into such where the indexes $\la$ and $\la'$ are attached to parts of the vector $p$, i.e. to $l_\la$, $d_\la$, $h_\la$ and the corresponding ones with the index $\la'$, and into such where the indexes are attached to the constant tensors, i.e., to $\delta^{||}_{\la\la'}$, $\delta^{\perp}_{\la\la'}$ and $F_{\la\la'}$. Let's call these parts the $p_\la p_{\la'}$-part and the $\delta_{\la\la'}$-part correspondingly. Second, from the $p_\la p_{\la'}$-part the first four form factors, $\Pi^{(i)}$ with $i=1,2,3,4$, can be restored uniquely. And third, from calculating $\Pi_{\la\la'}(p)p_{\la'}$, which can be done without the necessity to integrate by parts, the remaining two form factors can be restored uniquely.

For the restauration of the first four form factors from the $p_\la p_{\la'}$-part it is sufficient to consult the formulas \Ref{ts2} and \Ref{rep2}. The $p_\la p_{\la'}$-part results from the quantities $P_\la$ and $Q_\la$ whereas the $\delta_{\la\la'}$-part results in the coefficients $a$, $b$ and $c$ in \Ref{ts2}. Now from formula \Ref{rep2} it is seen that $a$, $b$ and $c$ do not enter the coefficients in front of the first four tensor stuctures, hence they do not enter the first four form factors.

For the restauration of the last two form factors we first calculate the non transversal part, $\Pi_{\la\la'}(p)p_{\la'}$. The result will be obtained in the form
\be\label{rhsc}\Pi_{\la\la'}(p)p_{\la'}=K_{\la\la'}(p) \Sigma_{\la'}(p).\ee
The fulfilment of Eq. \Ref{doubletransvp} follows from the transversality of $K_{\la\la'}(p)$. The factor $\Sigma_{\la'}(p)$ has the form
\be\label{Sigsig}\Sigma_{\la'}(p)=\sigma_l l_\la + \sigma_d id_\la +\sigma_h h_\la ,
\ee
where the $\sigma_i$ will be calculated explicitly, see subsection
\ref{sectnontrch}. It remains to derive the easy formulas
\bea\label{} K_{\la\la'}(p)l_{\la'}&=&T^{(5)}_{\la\la'}p_{\la'} \\
K_{\la\la'}(p)id_{\la'}&=&T^{(6)}_{\la\la'}p_{\la'} \\K_{\la\la'}(p)h_{\la'}&=&-T^{(5)}_{\la\la'}p_{\la'}
\eea
and we obtain for the r.h.s of Eq. \Ref{rhsc}
\be
\label{}K_{\la\la'}(p) \Sigma_{\la'}(p)=(\sigma_l-\sigma_h)T^{(5)}_{\la\la'}p_{\la'}+\sigma_d T^{(6)}_{\la\la'}p_{\la'}
\ee
If we multiply Eq.\Ref{expp} from the right by $p_{\la'}$ and compare with Eq.\Ref{rhsc} we identify the last two form factors,
\bea\label{T56}\Pi^{(5)}(p^2+2iF)&=&\sigma_l-\sigma_h , \nn \\
\Pi^{(6)}(p^2+2iF)&=&\sigma_d .
\eea
%

\subsection{Calculation of the Charged Polarization Tensor}\label{sectccpt}
\begin{figure}[htbp]\begin{picture}(10,3.5)
\put(2,0.19){ \psfig{height=3.0cm,file=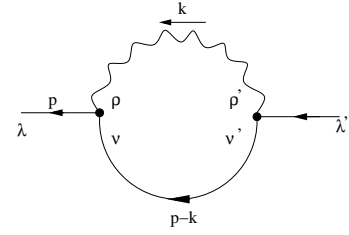}}
\end{picture}
\caption{The charged polarization tensor}
\label{figure:Picha}
\end{figure}

The charged polarization tensor has the following representation in momentum space (see Fig. \ref{figure:Picha})
\bea\label{CPT}&&
\Pi_{\la\la'}(p)=\Gamma_{\la\nu\rho}G_{\nu\nu'}(p-k)\Gamma_{\la'\nu'\rho'}G_{\rho\rho'}(k)\\&&
+(p-k)_{\la}G(p-k)k_{\la'}G(k)+k_{\la}G(p-k)(p-k)_{\la'}G(k)\nn ,
\eea
where the integration over the momentum $k$ is assumed.
The second line is the contribution from the ghosts. The  vertex factor is the same as \Ref{vfbg} but with renamed indices, it reads explicitly
\be\begin{array}{rclll}
\label{vfch}\Gamma_{\la\nu\rho}&=&
(k-2p)_\rho  \ g_{\la\nu}&+g_{\rho\nu}(p-2k)_\la&+g_{\rho\la}(p+k)_\nu , \\
&\equiv & \ \ \ \ \Q_\rho  \ \ \ \ \ g_{\la\nu}&+g_{\rho\nu} \ \ \ \ \Qb_\la&+g_{\rho\la} \ \ \ \ \Qt_\nu  .
\end{array}
\ee

In this section we use the parametric representation of the
propagators in the following form,
\be
\label{propc0}G(p-k) =\int_0^\infty ds \ e^{-s(p-k)^2} \mbox{ , }\quad  G(k)=\int_0^\infty dt \ e^{-tk^2}
\ee
for the scalar lines and
\bea\label{propc}
G_{\nu \nu'} (p-k)&=&\int_0^\infty ds \ e^{-s(p-k)^2}E_{\nu\nu'} \nn \\ 
G_{\rho \rho'} (k)&=&\int_0^\infty dt \ e^{-tk^2}g_{\rho\rho'} \eea
for the vector lines with
\be
\label{Ec}E_{\nu\nu'}=\delta^{||}_{\la\la'}-iF_{\la\la'}\sinh(2s)+\delta^\perp_{\la\la'} \cosh(2s)
\ee
which is the same as $E^{s}$ in the preceding section. Since we use the Feynman  gauge, $\xi=1$,  the propagator
of the neutral gluon is $G_{\rho\rho'}(k)=g_{\rho\rho'}\ G(k)$.

The momentum integration which is over $k$ here can be converted into the parametric integrals and  averaging in an auxiliary space again following \cite{Schwinger:1973kp}. The basic exponential reads
\be
\label{Thetac}\Theta=e^{-s(p-k)^2}e^{-tk^2}.
\ee
We denote the integration over the momentum $k$   by the average $\langle \dots\rangle$. It can be done delivering
\be \langle \Theta\rangle (h^2) =\frac{\exp\left[-\frac{st}{s+t}l^2-m(s,t)h^2\right]}{(4\pi)^2(s+t)\sqrt{N}} .
\label{Thetac1}
\ee
By the argument $h^2$ we indicated  the dependence of the average   on the external momentum. Of course it depends on $l^2$ too. Further, in Eq.\Ref{Thetac1} the  notations \\$N=\frac14\left((\szs+2t)^2-(\czs-1)^2\right)$ and
\bea\label{mst}
m(s,t)&\equiv&\left(s-{\rm arctanh}\frac{\czs-1}{\szs+2t}\right)
\nn \\ &=&\frac12 \ln\frac{1+2t-e^{-2s}}{1-(1-2t)e^{-2s}}
\eea
are introduced. In view of Eq.\Ref{p2+2iF} and  representation \Ref{expp} in terms of form factors it is meaningful to rewrite \Ref{Thetac1} as
\bea
\label{Thetac2}\langle \Theta\rangle (h^2) &=&\frac{\exp\left[-\frac{st}{s+t}l^2-m(s,t)(h^2+2iF)\right]}{(4\pi)^2(s+t)\sqrt{N}} \ {\cal Z}\nn\\&\equiv&\langle \Theta \rangle  (p^2+2iF) \ {\cal Z}
\eea
with
\be
\label{Z}{\cal Z}=-E^\top\frac{D}{D^\top} \ .
\ee
The quantities like $D$ and $A$ in the proceeding section appear
here too, but they have a different meaning,
\be
\label{A}A=E-1 \qquad \mbox{and} \qquad D=A-2itF.
\ee
Explicitly we need the following combinations,
\bea
\label{AD}
\frac{A}{-2iF}&=&s\delta^{||}-\frac{i}{2}F(\czs-1)+\frac12   \delta^\perp \szs   \nn \\
\frac{A}{D}&=&\frac{s}{s+t}\delta^{||}-iF \ \frac{t(\czs-1)}{2N}  \nn \\ && +\delta^\perp\frac{\czs-1+t\szs}{2N}, \nn \\
\label{FD}\frac{-2iF}{D}&=&\frac{1}{s+t}\delta^{||}
+iF \ \frac{\czs-1}{2N} \nn \\ &&
+\delta^\perp\frac{\szs+2t}{2N}
\eea
and
\bea
\label{ZE}{\cal Z}E&=&-\frac{D}{D^\top}=
\delta^{||}-iF \ \frac{(\czs-1)(\szs+2t)}{2N}\nn\\ && +\delta^\perp\frac{(\czs-1)^2+(\szs+2t)^2}{4N}.
\eea
We remind that all these relations are to be understood as matrix multiplication and that all these matrixes commute. The matrix $E$ has in addition the property $EE^\top=1$ where $E^\top$ is the transposed to $E$. In the formulas \Ref{Thetac2} till
\Ref{ZE} we dropped the indexes $\la$ and $\la'$.

The momentum $p$ does not commute with $\Theta$, instead
\be
\label{}p_\la \Theta=\Theta p(s)_\la
\ee
holds with
\be
\label{}p(s)_\la=(Ep)_\la-(Ak)_\la.
\ee
The averages of $k_\la$, i.e., the reductions of the momentum integration involving vectors to scalar integrations, read
\be
\label{}\langle \Theta(p^2)k_\la\rangle =\langle \Theta(p^2)\rangle \left(\frac{A}{D}p\right)_\la
\ee
and
\be
\label{}\langle \Theta(p^2)k_\la k_{\la'}\rangle =\langle \Theta(p^2)\rangle \left[\left(\frac{A}{D}p\right)_\la
\left(\frac{A}{D}p\right)_{\la'}+\left(\frac{iF}{D^\top}\right)_{\la\la'}\right].
\ee
However, for the calculation of the $p_\la p_{\la'}$-terms we do not need the last term in the r.h.s. Next we use formula \Ref{Thetac2} and get the rules
\bea\label{rules}\langle\Theta(p^2)k_\la\rangle &\to &\langle \Theta(p^2+2iF)\rangle \left({\cal Z}\frac{A}{D}p\right)_\la ,  \\
\langle \Theta(p^2)k_\la k_{\la'}\rangle &\to&\langle \Theta(p^2+2iF)\rangle \left({\cal Z} \frac{A}{D}p\right)_\la
\left(\frac{A}{D}p\right)_{\la'} .\nn
\eea
We apply these rules now to $\Pi_{\la\la'}(p)$, Eq. \Ref{CPT}, and with formulas \Ref{vfch} to \Ref{propc} we get
\be\label{PiM}\Pi_{\la\la'}(p)=\int_0^\infty ds\int_0^\infty dt \ \langle \Theta(p^2+2iF)\rangle M(s,t)_{\la\la'}
\ee
with
\be\label{Mc}M(s,t)_{\la\la'}=\sum_{i=1}^3 \ \left(  P^{(i)}_{\la} Q^{(i)}_{\la'} +
{Q^{(i)}}^\top_{\la} {P^{(i)}}^\top_{\la'} \right)
\ee
where we introduced the notations
\be\label{PQ}\begin{array}{rclrcl}
P^{(1)}_{\la}&=&
\left({\cal Z}\left( E\Q(s)+\Qb(s)\frac{tr E}{2} +E^\top \Qt(s)\right)p\right)_\la  , \nn \\
Q^{(1)}_{\la}&=&\left(\Qb p \right)_{\la} ,\nn \\
P^{(2)}_{\la}&=&\left(  {\cal Z} E \Qt p\right)_{\la} , \nn \\
Q^{(2)}_{\la}&=&\left(   \Q(s) p\right)_{\la}, \nn \\
P^{(3)}_{\la}&=&\left(  {\cal Z} (p(s)-k)\right)_{\la} ,\nn \\
Q^{(3)}_{\la}&=&   k _{\la} .\end{array}
\ee
Here $X(s)$, $Y(s)$ and $Z(s)$ are $X$, $Y$ and $Z$ with $p \ $ substituted by $p(s)$, accordingly.
We note that the relations ${\cal Z}\Q(s)=\Q^\top$, ${\cal Z}\Q=\Q^\top(s)$ hold. The same relations hold for $\Qb$ and $\Qt$.

In \Ref{Mc} and \Ref{PQ} the individual contributions are grouped as follows. From Eqs. \Ref{CPT} and \Ref{vfch} we get poducts involving two factors out from $X$, $Y$ and $Z$ in \Ref{vfch}. Denoting these symbolically for the moment e.g. by $XY$, and dropping all other factors we have to consider all possible products. Out of them some do not contribute to the $p_\la p_{\la'}$-terms, namely $XX$ and $ZZ$. From the remaining we got $P^{(1)}=X+\frac12 Y+Z$ and $Q^{(1)}=Y$ so that the product $P^{(1)}Q^{(1)}$ in \Ref{Mc} covers the products $XY$, half of $YY$ and $ZY$. The product ${Q^{(1)}}^\top {P^{(1)}}^\top$ in \Ref{Mc} covers $YX$, half of $YY$ and $YZ$. In the same way we get $P^{(2)}=Z$ and $Q^{(2)}=X$ and conjugated. Finally, $P^{(3)}$ and $Q^{(3)}$ result from the ghost contributions. We note that in deriving these formulas in the contributions $XZ$ and $ZX$ we changed the order of the two $p$'s appearing there. But their commutator results in a $\delta_{\la\la'}$-term.

Next we use formulas \Ref{FD} and \Ref{ZE} to get explicit
expressions for the $P$ and $Q$ in \Ref{PQ}. After some
calculation we get
\bea\label{PQc2}
P^{(1)}&=&
-\frac{s-t}{s+t} \czs \ \delta^{||}+ \left(3\szs-t\frac{\p}{N}\right) \ iF
\nn \\ &&
+\left(-1+t\frac{\q}{N}\right)\delta^\perp ,\nn \\
Q^{(1)}&=&
-\frac{s-t}{s+t}  \ \delta^{||}+ t\frac{\p}{N} \ iF
+\left(-1+t\frac{\q}{N}\right)\delta^\perp   , \nn \\
P^{(2)}&=&\frac{2s+t}{s+t}  \ \delta^{||} +\frac{\p(t-\q)}{N}  \ iF
+\left(\frac{\p^2+\q^2}{2N}-\frac{t\q}{2N}\right)\delta^\perp ,\nn \\
Q^{(2)}&=&
-\frac{s+2t}{s+t}  \ \delta^{||}
\nn \\ &&
+ \left(\frac{\alpha}{4N}-t\frac{\p\czs-\q\szs}{2N}\right) \ iF  ,\nn \\ &&
+\left(-\frac{\beta}{4N}+t\frac{\p\szs-\q\czs}{2N}\right)\delta^\perp    ,\nn \\
P^{(3)}&=&\frac{t}{s+t}\ \delta^{||}-\frac{t\p}{2N} \ iF+t\frac{t\q}{2N} \ \delta^\perp ,
\nn \\
Q^{(3)}&=&\frac{s}{s+t}\ \delta^{||}-\frac{t\p}{2N} \ iF+\left(1-\frac{t\q}{2N}\right)\delta^\perp
\eea
with
\bea\label{}\alpha&=&4t(\czs-1)+4t^2\szs, \nn \\
\beta&=& 2(\czs-1)+4t\szs+4t^2\czs \nn , \\
p&=&\czs-1 ,\nn \\
q&=&\szs+2t.
\eea
Now we remark that all expressions in \Ref{PQc2} have just that structure which was described in Eqs.\Ref{Pla} and \Ref{Qla} so that the corresponding $r$, $s$, $\alpha$, $\beta$, $\gamma$ and $\delta$ can be identified. Finally, we define a representation in parallel to \Ref{formfM} for the corresponding form factors,
\be\label{formfMc} \Pi^{(i)}(p)=\int_0^\infty\int_0^\infty \ ds \ dt \ \langle\Theta(p^2+2iF)\rangle \ M^{(i)}(s,t) \
\ee
and get
\bea\label{Pi14}
&& M^{(1)}(s,t)=4-2\left(\frac{s-t}{s+t}\right)^2 \czs  \nn \\
&& M^{(2)}(s,t)=
-2\left(-1+\frac{\q}{N}\right)^2-2\left(3-\frac{t\p}{N}\right)\frac{t\p}{N}\nn \\ &&
-2\left(\frac{\p^2+\q^2}{2N}-\frac{t\q}{2N}\right)
\left(\frac{-\beta}{4N}+t\frac{\p\szs-\q\czs}{2N}\right)\nn \\ &&
-2\left(-\frac{\p\q}{N}+\frac{t\p}{N}\right)
\left(\frac{\alpha}{4N}-t\frac{\p\czs-\q\szs}{2N}\right)
\nn \\ &&
-\frac{t\q}{N}\left(1-\frac{t\q}{2N}\right)+\left(\frac{t\p}{2N}\right)^2
\nn \\
&& M^{(3)}(s,t)=\frac{s-t}{s+t} \ \left(-1+\frac{t\q}{N}\right)(1+\czs) \nn \\ &&
-\frac{2s+t}{s+t}\left(\frac{-\beta}{2N}+t\frac{\p\szs-\q\czs}{2N}\right)\nn \\ &&
+\frac{s+2t}{s+t}\left(\frac{\p^2+\q^2}{2N}-\frac{t\q}{2N}\right)
-\frac{t}{s+t}\left(1-\frac{t\q}{2N}\right)-\frac{s}{s+t}\frac{t\q}{2N}
\nn \\
&& M^{(4)}(s,t)=\frac{s-t}{s+t}\left(3-\frac{t\p}{N}(1+\czs)\right)
\nn \\ &&
+\frac{2s+t}{s+t}\left(\frac{\alpha}{4N}-t\frac{\p\czs-\q\szs}{2N}\right)\nn \\ &&
+\frac{s+2t}{s+t}\left(-\frac{\p\q}{N}+\frac{t\p}{N}\right)
+\frac{s-t}{s+t} \frac{t\p}{2N}.
\eea
In this way the first four form factors are found in a representation that only the parametric integrals remain to be done.

For vanishing magnetic field which is equivalent to vanishing $s$ and $t$, in the lowest order we obtain the same result as for the neutral polarisation tensor (cf. Eq.\Ref{zeroth}),
\be\label{zerothch}M^{(i)}=4-2\left(\frac{s-t}{s+t}\right)^2 ~~~~ (i=1,2,3)\ee
and the $M^{(i)}$ with $i=4,5,6$ vanish (for $i=5,6$ see Eq.\Ref{Pi56}).

\subsection{The Non-Transversal Part of the Charged Polarization Tensor}\label{sectnontrch}
Here we make use of some transformation under the sign of the momentum integration. We start from the obvoious formula
\be\label{Gak2}\Gamma_{\la'\nu'\rho'}p_{\la'}=K_{\nu'\rho'}(p-k)-K_{\nu'\rho'}(k).
\ee
Further we use \Ref{obr} and \Ref{pGGp}.  As befor we drop all contributions in which  one of the two denominators cancels against a factor in the numerator. In the ghost contribution we use the identity $kp=\frac12(p^2+k^2-(p-k)^2)$. In this way we get in the next step
\bea\label{Pip}\Pi_{\la\la'}(p)p_{\la'}&=&
\Gamma_{\la\nu\rho}G_{\nu\nu'}(p-k)\left[K_{\nu'\rho}(p-k)-
K_{\nu'\rho}(k)\right]\frac{1}{k^2}    \nn \\ && \hspace{-1cm}
+(p-k)_{\la}\frac{1}{(p-k)^2}pk\frac{1}{k^2}
+k_{\la}\frac{1}{(p-k)^2}(p-k)k\frac{1}{k^2}    \nn \\
&=&-\Gamma_{\la\nu\rho}(p-k)_\nu\frac{1}{(p-k)^2}(p-k)_\rho\frac{1}{k^2}
  \\ && \hspace{-0.5cm}+\Gamma_{\la\nu\rho}k_\rho G_{\nu\nu'}(p-k)k_{\nu'}\frac{1}{k^2}
  +\frac12 p_{\la}\frac{1}{(p-k)^2}p^2\frac{1}{k^2}.\nn
\eea
Continuing with the mentioned manipulations we get
\bea\label{Pip1}
\Pi_{\la\la'}(p)p_{\la'}&=&
-\left[K_{\la\rho}(p)-K_{\la\rho}(k)\right]\frac{1}{(p-k)^2}(p-k)_{\rho}\frac{1}{k^2}
 \nn \\ &&
+\left[K_{\la\nu}(p-k)-K_{\la\nu}(p)\right] G_{\nu\nu'}(p-k)k_{\nu'}\frac{1}{k^2}
\nn \\ &&
 +\frac12 p_{\la}\frac{1}{(p-k)^2}p^2\frac{1}{k^2}
     \nn \\
 &&\hspace{-2cm}=
 -K_{\la\rho}(p)\frac{1}{(p-k)^2}(p-k)_{\rho}\frac{1}{k^2}
 -\frac{1}{(p-k)^2}k_\la (p-k)k\frac{1}{k^2}
 \nn \\ && ~
 \hspace{-2cm}-(p-k)_\la\frac{1}{(p-k)^2}(p-k)k\frac{1}{k^2}
 \nn \\ &&\hspace{-2cm} -K_{\la\nu}(p)G_{\nu\nu'}(p-k)k_{\nu'}\frac{1}{k^2}
+\frac12 p_{\la}\frac{1}{(p-k)^2}p^2\frac{1}{k^2}  .\eea
Now the ghost contribution cancels against the second and the
third terms. Finally we obtain
\be\label{Pip4}\Pi_{\la\la'}(p)p_{\la'}=K_{\la\rho}(p) \ \Sigma_\rho(p)
\ee
with
\be\label{Sigla}\Sigma_\rho(p)=\left(\frac{-1}{(p-k)^2}(p-k)_\rho-
G_{\rho\nu'}(p-k)k_{\nu'}\right)\frac{1}{k^2}.
\ee
From \Ref{Pip4} it is obvious that $p_\la
\Pi_{\la\la'}(p)p_{\la'}=0$ holds. Now we calculate the r.h.s. of
Eq.\Ref{Sigla}. Using the same formulas as in sect. \ref{sectccpt}
we get
\be\label{}\Sigma_\la(p)=\int_0^\infty ds\int_0^\infty dt \ \left(\langle \Theta(p^2+2iF)\rangle {\cal Z}\left(-(p-k)-Ek\right)\right)_\la .
\ee
Using formulas \Ref{FD} and \Ref{ZE} we transform
\be\label{}{\cal Z}\left(-(p-k)-Ek\right)=
{\cal Z}\left(-1-\frac{A^2}{D}\right)p
\ee
and with \Ref{A} we come to
\bea\label{}{\cal Z}\left(-1-\frac{A^2}{D}\right)&=&
\frac{E^\top}{D^\top}(A-2itF+A^2)  \\
&=&-\left(\frac{A}{-2iF}+tE^\top\right)\left(\frac{-2iF}{D}\right)^\top .\nn
\eea
Now we use the explicit expressions from \Ref{FD} and  obtain
\bea\label{}&&{\cal Z}\left(-1-\frac{A^2}{D}\right)\nn \\ &&
-\delta^{||} \nn \\ &&
+iF\frac{1}{4N} \Big((\czs-1)(\szs+2t\czs)\nn \\ && -(\szs+2t)(-\czs+1+2t\szs)\Big)\nn \\ &&
+\delta^\perp \frac{1}{4N}\Big((\czs-1)(-\czs+1+2t\szs)\nn \\ &&-(\szs+2t)(\szs+2t\czs)\Big) .
\eea
Further we multiply by the vector $p$ and with \Ref{Sigsig} we identify
\bea\label{}
\sigma_l&=&\int_0^\infty ds\int_0^\infty dt \ \langle \Theta(p^2+2iF)\rangle \sigma_l(s,t),\nn\\
\sigma_d&=&\int_0^\infty ds\int_0^\infty dt \ \langle \Theta(p^2+2iF)\rangle \sigma_d(s,t),\nn\\
\sigma_h&=&\int_0^\infty ds\int_0^\infty dt \ \langle \Theta(p^2+2iF)\rangle \sigma_h(s,t)
\eea
with
\bea\label{}
&&\sigma_l(s,t)  -1 ,\nn \\
&&\sigma_d(s,t)=  \frac{\szs(2t^2-\czs+1)}{2N} , \nn \\
&&\sigma_h(s,t)= \\ &&\frac{\czs(\czs-1)+2t\szs+2t^2\czs)}{2N},\nn
\eea
where some simplifications had been made.
Finally we get with \Ref{T56} the last two form factors in the representation \Ref{formfMc},
\bea\label{Pi56}&&
M^{(5)}(s,t)=\nn \\ &&1-\frac{\czs(\czs-1)+2t\szs+2t^2\czs)}{2N},\nn \\
&&M^{(6)}(s,t)= \frac{\szs(2t^2-\czs+1)}{2N}.
\eea
By these formulas the calculation of the polarization tensor of the charged gluons is fi\-nished.
\section{Renormalization of the Charged Polarization Tensor}\label{sectrench}
In order to discuss the renormalization of the charged polarization tensor we change the notations and represent the form factors which are given by Eq.\Ref{PiM} and Eq.\Ref{formfMc} as
\be\label{Pir}\Pi^{(i)}(p)=\int_0^\infty ds\int_0^\infty dt  \ \frac{{M}^{(i)}(s,t)}{(s+t)\sqrt{N}}\ e^{-H(s,t)}
\ee
with
\be\label{Ht}H(s,t)=\frac{st}{s+t} \ l^2+m(s,t)\ (h^2+2iF),
\ee
where the ${M}^{(i)}(s,t)$ are given by Eqs. \Ref{Pi14} and \Ref{Pi56} and $m(s,t)$ by Eq.\Ref{mst}. We change variables according to $s=\la x$, $t=\la(1-x)$ and rewrite Eq.\Ref{Pir} as
\be\label{Pir2}\Pi^{(i)}(p)=\int_0^\infty \frac{d\la}{\la} \ f_i(\la)
\ee
with
\be\label{fla}f_i(\la)=\int_0^1dx \ \frac{\la{M}^{(i)}(s,t)}{\sqrt{N}}\ e^{-H(s,t)}.
\ee
For convenientness of representation we keep the variables $s$ and $t$ although they are to be substituted by $\la$ and $x$.

Now we restore the dependence on the magnetic field $B$. This can be done by substituting $\la\to \la B$ and $l^2\to l^2/B$ in $f_i(\la)$. We remind that $\Pi^{(i)}(p)$ is dimensionless. Further we use the representation in Minowski space which can be obained from the Euclidean representation which we used so far by rotating 'back', i.e., by substituting $\la\to -i\la$ and adding a '$-\lambda\epsilon $' in the exponential defining the causal propagator in the standard way with $\epsilon>0$. The momentum becomes $l^2\to -l_0^2+l_3^2$. Further we introduce an ultraviolet regularization (with regularization parameter $\delta>0$ and arbitrary massive parameter $\mu$) and come to the representation
\be\label{Pil}\Pi^{(i)\rm reg}(p)=\int_0^\infty\frac{d\la}{\la} \ (\la \mu^2)^\delta \  f_i(-i\la B) \ e^{-\la\epsilon}.
\ee
The ultraviolet divergences come from the small $\la$. We note that in the limit of $\la\to0$  the $f_i(\la)$ for $i=1,2,3$ become just that for a polarization tensor without magnetic field and for $i=4,5,6$ they vanish (see Eqs.\Ref{zeroth} and \Ref{zerothch}). Taking into account \Ref{sumTipa} in this way the renormalization can be done indeed by subtracting all contributions which do not depend on the magnetic field.
By adding and subtracting $f_i(0)$ they can be separated,
\be\label{Pidiv}\Pi^{(i)\rm div}(p)=\left(\frac{1}{\delta}+\ln \mu^2\right)f_i(0)
\ee
and the renormalized form factors are
\bea\label{Piren}&&\Pi^{(i)\rm ren}(p)=\ln \la_0 \ f_i(0)+\int_0^{\la_0} \frac{d\la}{\la} \ (f_i(-i\la B )-f_i(0))\nn \\ &&+\int_{\la_0}^\infty \frac{d\la}{\la} \ f_i(-i\la B) \ e^{-\la\ep},
\eea
where we divided the integration region at an arbitrary point $\la_0$. Obviously \Ref{Piren} does not depend on $\la_0$.

In this way we obtained the renormalized form factors and, strictly speaking, the final expression. It is finite and can be calculated numerically. But it is useful to give an numerical example for that which, for instance, demonstrates the appearance of the imaginary part resulting from the tachyonic mode. The most typical example one can think of is to take the polarization tensor averaged in the unperturbed tree level states \Ref{chstates}. This gives the first perturbative correction,
\be\label{corr}
\Delta E^2_n=\langle p_{||},n,s\mid_\alpha \Pi_{\alpha\beta}(p) \mid p_{||},n,s\rangle_\beta ,
\ee
to the tree level energy  $E^2_{\mbox{tree}}=p_3^2+B(2n+1)$ (cf. \Ref{treelevel}). Taking into account the the matrix elements \Ref{maels} in the physical states $s=1,2$ from representation \Ref{expp} of the polarization tensor we get
\bea\label{avs}&&
\Delta E^2_n(s=1)\nn \\ &&=(\Pi^{(2)} -\Pi^{(3)}+ \Pi^{(5)} )(2n+1)- \Pi^{(4)}+\Pi^{(6)}, \nn \\
&&\Delta E^2_n(s=2) \\ &&=(-\Pi^{(1)} +\Pi^{(3)}+ \Pi^{(5)})(2n+1)- \Pi^{(4)}+\Pi^{(6)}.\nn
\eea
where the on-shell condition $-l_4^2=l_0^2=p_3^2+B(2n+1)$ was used.

It must be noticed, however, that the form factors $\Pi^{(i)}(p_{||}^2,p_\perp^2)$ have on-shell a logarithmic infrared singularity which is typical for a graph with massless particles taken on-shell. We regularize this divergence by an auxiliary gluon mass $m_g$ and perform the calculation for $m_g\to0$. This mass is, of course, the same as the parameter $\epsilon$ in Eq.\Ref{Piren} so that we substitute $\epsilon\to im_g^2$ there. Now the the form factors depend only on the magnetic field, on the auxiliary mass and on the   number $n$ of the states in \Ref{corr} and the calculation  is reduced to a numerical evaluation of the double parametric integrals in \Ref{fla} and \Ref{Piren}. However, in this form the integrand is an oscillating function and not very friendly for numerical evaluation. It would be useful to perform the Wick rotation. It is known that this is impossible in a direct way because  the euclidean integral diverges exponentially due to the tachyonic mode. We demonstrate a way to overcome this problem. We start from representation \Ref{Piren} taken on-shell with the auxiliary mass $\epsilon\to im_g^2$. After that we rescale the variable $\la$ by $\la\to \la/B$ and make the choice $\la_0=1/B$. After that the magnetic field enters only through the momentum  in the combination $l^2/B$ and $h^2/B$ in $H(s,t)$, \Ref{Ht}. In Minkowski space the momentum is $l^2=-l_0^2+l_3^2$ and on-shell we have $l_0^2=l_3^2+B(2n+1)$ and $h^2=B(2n+1)$ so that with $l^2/B\to (2n+1)$  the $B$-dependence drops out leaving behind only the dependence on the number $n$ of the Landau level. Of course, now the first term in the r.h.s. of Eq. \Ref{Piren} contains $\ln B$.

The tachyonic mode manifests itself in the behavior of the functions $M^{(i)}(s,t)$ for large $s$. It holds
\be\label{mi}  M^{(i)}(s,t) =e^{2s} m_i(s,t)+O\left(1\right)
\ee
for large $s$ and the factor $e^{2s}$ determines the leading behavior of the whole integrand.
The functions $m_i(s,t)$ read
\be\label{mist} \begin{array}{rclrclrcl}
m_1(s,t)&=&-\left(\frac{s-t}{s+t}\right), &
m_2(s,t)&=&\frac{2(1+t)}{1+2t},\\ [6pt]
m_3(s,t)&=&\frac{s+5t}{2(s+t)(1+2t)}, & m_4(s,t)&=&\frac{s+4st-t(7+4t)}{2(s+t)(1+2t)}, \\ [6pt]
m_5(s,t)&=&\frac{1}{2(1+2t)}, &
m_6(s,t)&=&\frac{-1}{2(1+2t)}.
\end{array}\ee
We divide the whole expression for the form factor into what comes from \Ref{mi} (part A) and the rest (part B),
\bea\label{fAB}f^A_i(\la)&=&\int_0^1 dx \ e^{2s} m_i(s,t) \frac{\la}{\sqrt{N}}\ e^{-H(s,t)}, \\
f^B_i(\la)&=&\int_0^1 dx \  \left( M^{(i)}(s,t)-e^{2s} m_i(s,t)\right) \frac{\la}{\sqrt{N}}   \ e^{-H(s,t)}  . \nn
\eea
Further we remark that on-shell the function in the exponential is
\bea\label{}H(s,t)&=&\left(-\frac{st}{s+t}+m(s,t)\right)(2n+1),\nn\\&\equiv&-h(s,t)(2n+1).
\eea
The function $h(s,t)$ has the properties
\bea\label{}h(-s,-t)&=&-h(s,t),\nn \\
0&\le &h(s,t)\le\frac14 \la \ \ \ \ (s,t\ge 0),\nn \\
h(\la x,\la(1-x))&=&\frac13\la^3x^4+O(x^5),
\eea
from which it follows that for $n=-1$  the leading behavior of $f^A_i(\la)$ for $\la\to\infty$ is indeed determined by the factor $e^{2s}$. For higher $n$ one had to include more orders in the expansion \Ref{mi} so that we restrict ourselves to $n=-1$ in the following.

In this way the usual Wick rotation is impossible to do in part A. This is in line with the picture in momentum space. Using the causal propagator we have the usual poles below the positive real $k_0$-axis and above the negative one. In addition there are poles from the tachyonic mode which are located on the opposite side of the imaginary axis. In these contributions the Wick rotation would deliver additional contributions resulting from these poles. But it is possible to rotate the contour clockwise, i.e., in the opposite direction, $\la\to-i\la$. We do this in part A. Of course, in part B we rotate in the usual direction, $\la\to i\la$.

We are going to do these rotations in formula \Ref{Piren}. However, there the integration region is divided into two parts with different functions. To get one integral we integrate by parts and obtain
\bea\label{Piren1}&&\Pi^{(i)\rm ren}(p)=-\ln B \ f_i(0)  \\ &&-\int_0^\infty  d\la \ \ln \la \ \frac{d}{d\la}\left(f^A_i(-i\la )+f_i^B(-i\la )\right) \ e^{i\la m_g^2}.\nn
\eea
The $f_i(0)$ follow from Eq.\Ref{zerothch} after integration over $x$, their values are
\be\label{fifree}f_i(0)=\frac{10}{3}  \ \mbox{for}  \ i=1,2,3  \quad \mbox{and} \quad f_i(0)=0 \ \mbox{for}  \ i=4,5,6.\ee

Now we perform the Wick rotation in part B and the opposite one in part A. From $\ln \la$ the additional contributions
\be\label{}i\frac{\pi}{2}\left(f^B_i(0)-f^A_i(0)\right)=
i\frac{\pi}{2}\left\{\begin{array}{cl}10/3 & (i=1,2),\\
1/3 & (i=3),\\3&(i=4),\\0&(i=5,6)\end{array}\right.
\ee
follow which contribute to the imainary part of the form factors.
In the remaining expressions of part A the rotation results in the substitution $\la\to-\la+i0$ in the Euclidean formulas. The prescription $+i0$ is necessary because the functions $m_i(s,t)$, Eq.\Ref{mist}, have for $i=2,\dots,6$ a pole in $t=-1/2$. This pol delivers a further contribution to the imaginary part.

Finally we integrate by parts back and obtain
\be\label{Piren2}
\Pi^{(i)\rm ren}(p)=-\ln B \ f_i(0)+i\frac{\pi}{2}\left(f^B_i(0) -f^A_i(0)\right)+\Pi_{(i)}^A+\Pi_{(i)}^B
\ee
with
\bea \label{PiAB}
\Pi_{(i)}^A&=&
\int_0^1\frac{d\la}{\la} \
\left(f^A_i(-\la )-f^A_i(0)\right)+\int_1^\infty \frac{d\la}{\la} \ f^A_i(-\la )  \nn \\
\Pi_{(i)}^B&=&\int_0^1\frac{d\la}{\la} \
\left(f_i^B(\la )-f_i^B(0)\right)+
\int_1^\infty \frac{d\la}{\la}  \ f_i^B(\la ) \ e^{i\la m_g^2}.\nn \\
\eea
Now we have to deal with the infrared divergence for $m_g\to0$. It appears only from $f^B_i(\la)$  because of
\be\label{as25}f^B_i(\la) \to \frac12 \quad \mbox{for} \quad \la\to\infty \quad i=2,\dots,6
\ee
holds. By adding and subtracting this asymptotic value we get
\bea\label{26}
\Pi_{(i)}^B&=&\int_0^1\frac{d\la}{\la} \
\left(f_i^B(\la )-f_i^B(0)\right) +
\frac12\left(-\ln\frac{m_g^2}{B}+C\right)\nn \\ &&+\nn
\int_1^\infty \frac{d\la}{\la}  \left( f_i^B(\la )-\frac12\right)+O\left(m_g^2\right)\\
&\equiv& \frac12\left(-\ln\frac{m_g^2}{B}+C\right)+\tilde{\Pi}^{B}
\eea
for $i=2,\dots,6$, where $C$ is Euler's constant. For $i=1$, $f_i^B(\la )$ vanishes for $\la\to\infty$ and $\Pi_{(1)}^B $ is given by Eq.\Ref{PiAB} for $m_g=0$ directly in the same way as $ \Pi_{(i)}^A$ since the $f^A_i(-\la )$ vanish too.

Now the integrations can be done and deliver numbers, see Table \ref{table}. In this way we get the correction
\bea\label{corr}
\Delta E^2_n&=&\langle p_{||},n,s\mid_\alpha \Pi_{\alpha\beta}(p) \mid p_{||},n,s\rangle_\beta \\
&=&\left(-\frac12\left(-\ln\frac{m_g^2}{B}+C\right)+1.807 - 7.15 i \right)\frac{g^2}{16\pi^2}B \nn
\eea
to the energy of the lowest level. Here we restored the dependence on all prefactors.

\begin{table}
  \centering
\begin{tabular}{ccccc}
  i & $\Pi^{A}$ & $\Pi^{B}$ & $\tilde{\Pi}^{B}$ \\\hline
  1 & 0.012 - 0.52i  & 6.16  &      \\
  2 & 0.62 + 1.90i     &   & -0.167    \\
  3 & 0.32 - 0.11i   &   & 4.69     \\
  4 & 0.18 + 0.11i  &   & 3.53     \\
  5 & 0.16 + 0.16i  &   &  -0.97    \\
  6 & -0.16 - 0.16i  &   &  0.31    \\

\end{tabular}

  \caption{Results of numerical integration}\label{table}
\end{table}


\section{Conclusions}\label{sectconcl}
To summarize the results pesented in this paper we note that the gluon polarization tensor in a homogeneous magnetic background field is not transversal. As a consequence the polarization tensor has 10 independent tensor structures and, accordingly, 10 form factors. Seven of these tensor structures are transversal and three are not. We found this non transversality from the explicit calulations in one loop order.
It shoud be mentioned that the non transversality  takes place in general. Only in special cases parts of the polarization tensor may be  transversal.  For example, this is known   for the neutral component in Fujikawa gauge.

For the calculation of the form factors Schwinger's proper time method was used. The form factors are represented as double parameter integrals, the integrands have a form close to what is known from similar calculations in QED. The are even not much more complicated. In general, in this representation an integration by parts in the parameter integrals is desirable in order to get all contributions in the same representation. In the case of the neutral polarization tensor this is simple, for the charged polarization tensor it is not known how to do that. We solved this problem by making use of a sufficiently explicit representation we derived for the non transversal part.

One of the  opened problems is the definition of suitable operator for the spin projection like $\sigma_{\mu\nu}$ in the spinor case. What can be done with the tensor structures found here is to investigate the level shifts and splitting in the magnetic field due to the radiation corrections given by the polarization tensor.

The tensor structures found in this paper may be used in two ways. First, they may serve as a input for the structures appearing in a Schwinger-Dyson equation in the magnetic background when the technique of the 2Pi functionals is used. Second, they may be generalized to include temperature which should be a quite straight forward task now.  Another interesting question is to investigate explicitly the dependence on the gauge parameter $\xi$ of, for example, the level shift and splitting or the gluon condensate which should turn out to be gauge invariant. Using the  methods developed here it should be  possible to calculate the $\xi$-dependent part for the charged component too, which is however, beyond the limit of this paper.

The same structures can also be used if the W-boson mass operator in a magnetic field is investigated. This issue will be discussed elsewhere.

Another interesting question of technical character is the integration by parts in the parameter integrals. For the neutral component it is known and serves as a check for the formulas obtained  insofar as from the direct calculation of the nontransversal part the same result must appear. In the neutral case this is indeed the case. But for the charged component this check is missing.

Another task left for future work is to use the Slavnov-Taylor identities (as given, for example, in \cite{Faddeev1980as}) to derive the nontransversal parts and to compare with \Ref{Pik4}, \Ref{Sigmu}, \Ref{Pip4} and \Ref{Sigla} which were derived from direct calculations.
%
\section*{Acknowledgement}
The authors thank A.Strelchenko for many helpful discussions and A.Shabad for consultation on the tensor structure of the polarization tensor. One of us (V.S.) was supported by DFG under grant number 436UKR17/19/03.  Also he thanks the Institute for Theoretical Physics of Leipzig University for kind hospitality.

\section*{Appendix A}\label{appx1}\renewcommand{\theequation}{A\arabic{equation}}\setcounter{equation}{0}
Here we note the explicit form of the operators \Ref{Tnp} after rotation according to \Ref{Bma},
\be\label{Tma}T^{(i)}_{\alpha\beta}=B^*_{\alpha\mu}T^{(i)}_{\mu\nu}B_{\nu\beta} .
\ee
They read
\be \begin{array}{ll}
T^{(1)}=\left(\begin{array}{cc}0&0\\0&\left(
\begin{array}{cc}l_4^2&-l_3l_4\\-l_3l_4&l_3^2\end{array}\right)\end{array}\right) ,\\ [25pt]
T^{(2)}=\left(\begin{array}{cc}\left(\begin{array}{cc}n-1&{a^\dagger}^2\\a^2&n+2\end{array}\right)&0\\0&0\end{array}\right) ,\\ [25pt]
T^{(3)}=\left(\begin{array}{cccc}l^2&0&-ia^\dagger l_3 & -ia^\dagger l_4\\
0&l^2&ial_3&ial_4\\
ial_3&-ia^\dagger l_3&2\hat{n}+1&0\\ial_4&-ia^\dagger l_4&0&2\hat{n}+1
\end{array}\right), \\ [25pt]
T^{(4)}=\left(\begin{array}{cccc}-l^2&0&ia^\dagger l_3 & ia^\dagger l_4\\
0&l^2&ial_3&ial_4\\
-ial_3&-ia^\dagger l_3&-1&0\\-ial_4&-ia^\dagger l_4&0&-1
\end{array}\right),\\ [30pt]
T^{(5)}=\left(\begin{array}{cccc}-l^2&0&0&0\\0&-l^2&0&0\\0&0&2\hat{n}+1&0\\0&0&0&2
\hat{n}+1\end{array}\right),&
 \\  [30pt] T^{(6)}=\left(\begin{array}{cccc}-2(\hat{n}-1)-l^2&0&0&0\\0&2(\hat{n}+2)
 +l^2&0&0\\0&0&1&0\\0&0&0&1\end{array}\right). \end{array}
\ee
The letter $\hat{n}$ in these formulas is the number operator, $\hat{n}=\frac{1}{2}(aa^\dagger+a^\dagger a)$.

The matrix elements of the  tensors can be calculated easily. For $i=1,2,3$ they follow from \Ref{evs}. The complete list for the transversal states reads
\be\label{maels}
\begin{array}{rclrcl}
\langle 1\mid T_1 \mid 1 \rangle &=&  0,    &\langle 2\mid T_1 \mid 2 \rangle &=& \frac{l_4^2h^2}{k^2},\\ [8pt]
\langle 1\mid T_2 \mid 1 \rangle &=& h^2 ,  &\langle 2\mid T_2 \mid 2 \rangle &=& 0,\\ [8pt]
\langle 1\mid T_3 \mid 1 \rangle &=& l^2 ,  &\langle 2\mid T_3 \mid 2 \rangle &=& k^2+\frac{l_3^2l_4^2}{k^2},\\ [8pt]
\langle 1\mid T_4 \mid 1 \rangle &=& \frac{l^2}{h^2} ,&\langle 2\mid T_4 \mid 2 \rangle &=& -\frac{k^4+l_3^2l_4^2}{h^2k^2},\\ [8pt]
\langle 1\mid T_5 \mid 1 \rangle &=& -l^2 , &\langle 2\mid T_5 \mid 2 \rangle &=& \frac{h^4-l_3^3l^2}{k^2},\\ [8pt]
\langle 1\mid T_6 \mid 1 \rangle &=& \frac{l^2+2h^2}{h^2} ,&\langle 2\mid T_6\mid 2 \rangle &=& \frac{h^4-l_3^3l_4^2}{k^2h^2}.
\end{array}
\ee


\begin{thebibliography}{10}

\bibitem{Agasian:2003yw}
N.~O. Agasian.
\newblock Phys. Lett. {\bf B562}, 257 (2003)

\bibitem{Skalozub:2002da}
V.~V. Skalozub, A.~V. Strelchenko.
\newblock Eur. Phys. J. {\bf C33}, 105 (2004)

\bibitem{Skalozub:2004ab}
V.~V. Skalozub, A.~V. Strelchenko.
\newblock Eur. Phys. J. {\bf C40}, 121 (2005)

\bibitem{Pollock}
M.~D. Pollock.
\newblock Int. J. Mod. Phys. {\bf D12}, 1289 (2003)

\bibitem{Kalashnikov:1982sc}
O.~K. Kalashnikov.
\newblock Fortschr. Phys. {\bf 32}, 525 (1984)

\bibitem{Enqvist:1994rm}
K.~Enqvist, P.~Olesen.
\newblock Phys. Lett. {\bf B329}, 195 (1994)

\bibitem{Starinets:1994vi}
A.~O. Starinets, A.~S. Vshivtsev, V.~C. Zhukovsky.
\newblock Phys. Lett. {\bf B322}, 403 (1994)

\bibitem{Elmfors:1998dr}
P.~Elmfors, D.~Persson.
\newblock Nucl. Phys. {\bf B538}, 309 (1999)

\bibitem{Skalozub:2000ay}
V.~V. Skalozub, A.~V. Strelchenko.
\newblock Phys. Atom. Nucl. {\bf 63}, 1956 (2000)

\bibitem{Nielsen:1978rm}
N.~K. Nielsen, P.~Olesen.
\newblock Nucl. Phys. {\bf B144}, 376 (1978)

\bibitem{Borisov:1984di}
A.~V. Borisov, V.~C. Zhukovsky, M.~Y. Knizhnikov.
\newblock Yad. Fiz. {\bf 39}, 1504 (1984)

\bibitem{Kaiser:1987cc}
H.~J. Kaiser, E.~Wieczorek, K.~Scharnhorst.
\newblock Phys. Appl. {\bf 14}, 239 (1987)

\bibitem{Schwinger:1973kp}
J.~S. Schwinger.
\newblock Phys. Rev. {\bf D7}, 1696 (1973)

\bibitem{Faddeev1980as}
A.~Slavanov, L.~Faddeev.
\newblock {\em {Gauge fields, Introduction to quantum theory}\/}
  (Benjamin/Cummings, Advanced Book Program, 1980)

\end{thebibliography}

\end{document}